\documentclass[reprint,pra,aps,amsmath,amssymb, twocolumn,superscriptaddress, raggedbottom]{revtex4-2}

\usepackage{subcaption}
\usepackage{graphicx}
\usepackage{dcolumn}   
\usepackage{bm}        
\usepackage{bbm}
\usepackage[colorlinks=true, 
urlcolor=blue,citecolor=blue,anchorcolor=blue,breaklinks=true]{hyperref}
\usepackage[capitalise]{cleveref}
\usepackage{dsfont}
\usepackage{mathtools}
\usepackage{amsthm, thmtools}
\usepackage{amsmath}
\usepackage{empheq}
\usepackage[ruled]{algorithm2e}
\usepackage{xcolor}
\usepackage[normalem]{ulem} 

\newcommand{\microspace}{\mspace{0.5mu}}

\def\<{\langle}
\def\>{\rangle}
\def \lket {\left|}
\def \rket {\right\rangle}
\def \lbra {\left\langle}
\def \rbra {\right|}
\newcommand{\ket}[1]{\lket\microspace #1 \microspace\rket}

\newcommand{\bra}[1]{\lbra\microspace #1 \microspace\rbra}

\newcommand{\avg}[1]{\left\langle#1\right\rangle}
\usepackage{mathtools}

\usepackage{bm}
\let\vec\bm 

\begin{document}
\title{Quantum DPLL and Generalized Constraints in Iterative Quantum Algorithms}

\author{Lucas~T.~Brady}
\email{Lucas.T.Brady@nasa.gov}
\affiliation{Quantum Artificial Intelligence Laboratory, NASA Ames Research Center, Moffett Field, California 94035, USA}

\author{Stuart Hadfield}
\affiliation{Quantum Artificial Intelligence Laboratory, NASA Ames Research Center, Moffett Field, California 94035, USA}
\affiliation{USRA Research Institute for Advanced Computer Science (RIACS), Mountain View, CA 94043, USA}

\date{\today}
\begin{abstract}
Too often, quantum computer scientists seek to create new algorithms entirely fresh from new cloth when there are extensive and optimized classical algorithms that can be generalized wholesale.  At the same time, one may seek to maintain classical advantages of performance and runtime bounds, while enabling potential quantum improvement.  Hybrid quantum algorithms tap into this potential, and here we explore a class of hybrid quantum algorithms called Iterative Quantum Algorithms (IQA) that are closely related to classical greedy or local search algorithms, employing a structure where the quantum computer provides information that leads to a simplified problem for future iterations.  Specifically, we extend these algorithms beyond past results that considered primarily quadratic problems to arbitrary $k$-local Hamiltonians, proposing a general framework that incorporates logical inference in a fundamental way.  As an application we develop a hybrid quantum version of the well-known classical Davis-Putnam-Logemann–Loveland (DPLL) algorithm for satisfiability problems, which embeds IQAs within a complete backtracking-based tree search framework.  Our results also provide a general framework for handling problems with hard constraints in IQAs.  We further show limiting cases of the algorithms where they reduce to classical algorithms, and provide evidence for regimes of quantum improvement.
\end{abstract}

\maketitle

\section{Introduction}

While some quantum algorithms offer the possibility of scaling advantage over classical algorithms, these algorithms are often relatively simple in their construction and lack much of the fine-tuning and sophistication that corresponding classical algorithms have accrued over decades of empirical testing and refinement.  Some of this sophistication will come to quantum algorithms with time, especially as they are implemented and used on real hardware.  An alternate method for jump-starting this sophistication is to directly adapt successful techniques and methods from classical protocols to novel hybrid quantum-classical algorithms.  
At the same time, such algorithms should be designed in a way that the quantum component is providing verifiable improvements, not being drowned out by the performance of the classical sophistication alone.

Classical optimization algorithms, especially, have had 
much 
time to ferment, with a body of literature developing rich flavors from advanced algorithms and competitions for the best heuristics.  
Many families of successful heuristics have benefited from a deploy-and-refine model, including iterative improvements based on empirical testing, cross-pollination across algorithms, as well as tailoring to specific problem subclasses. 
This work does not seek to dive directly into this pool of hyper-optimized heuristics but instead show how quantum subroutines can be methodically incorporated into existing algorithms, and analyzed.  To that end, we expand and formalize a class of hybrid quantum approaches called Iterative Quantum Algorithms \cite{Bravyi2020,dupont2023quantum,Brady2023,Finzgar2023} which rely on quantum subroutines mixed with classical computing.  Specifically this class of algorithms relies on three main steps at each iteration: a preparation step which repeatedly prepares and measures some quantum state, a selection step which picks some piece of problem information based on the preparation step measurement data, and a reduction step that then reduces the problem based off the selection.  
This new subproblem is sent through the same steps until the problem has become sufficiently reduced that it can be solved exactly, and that solution can be unwound using the sequence of selections to yield an exact or approximate solution to the original problem.

Quantum Optimization has traditionally relied on adiabatic-like procedures \cite{Kadowaki_1998,farhi2000,Crosson_2021} or parameterized quantum circuit ans\"atze, such as the Quantum Approximate Optimization Algorithm (QAOA)~\cite{Farhi2014,hadfield2019quantum} or the Variational Quantum Eigensolver (VQE)  \cite{Peruzzo2014,Cerezo2021} to reach the ground states of Hamiltonians that correspond to the optimization problem.  
For combinatorial optimization problems of interest, obtaining the ground state (i.e., optimal solution) is often NP-hard, and so we often must settle for the best approximate solution possible.
While these methods work and can even have analytic performance or runtime bounds in some cases, they are limited in that moving beyond relatively simpler unconstrained binary problems comes with considerable challenges or resource overhead. 
For instance, usually constraints are implemented as a penalty function added to the cost Hamiltonian that act as additional soft constraints, and while there are methods that can implement constraints in a hard manner, often through symmetry protection \cite{hadfield2019quantum}, in practice, these methods are potentially costly to implement and may be insufficient to preserve constraints in noisy environments~\cite{Maciejewski2024}.

Iterative quantum algorithms were first introduced in the context of MaxCut problems, equivalent to finding the ground state of Ising models or quadratic unconstrained binary optimization, under the name Recursive QAOA (RQAOA) \cite{Bravyi2020,Bravyi2022hybridquantum}.  This first implementation was a simple method for enhancing the performance of QAOA, especially for low depth algorithms.  The algorithm works by running the quantum algorithm, picking the edge with the highest absolute-value correlator, and fixing that correlation before sending the problem back for additional rounds with the quantum algorithm.  It is known that training variational quantum algorithms can be NP-Hard \cite{Bittel2021} and suffers from issues such as barren plateaus in the training landscape in the limit of large circuit depths \cite{Ragone_2024,Fontana_2024}.  Some approaches have sought to reduce the dependence on variational parameters by reusing parameters \cite{Akshay_2021,Wurtz_2021,Shaydulin_2023} or by control theory \cite{Magann_2022,Brady_2024}.  RQAOA sought to circumvent these hard training problems, as well as the performance limitations of low-depth or near-term quantum circuits, by relying on the quantum algorithm to decide the somewhat simpler task of which selection and reduction to make at each iteration, relaxing the complexities of solving the input problem directly and potentially alleviating the need for larger circuit depths.

\begin{figure}[t]
\includegraphics[width = 0.48\textwidth]{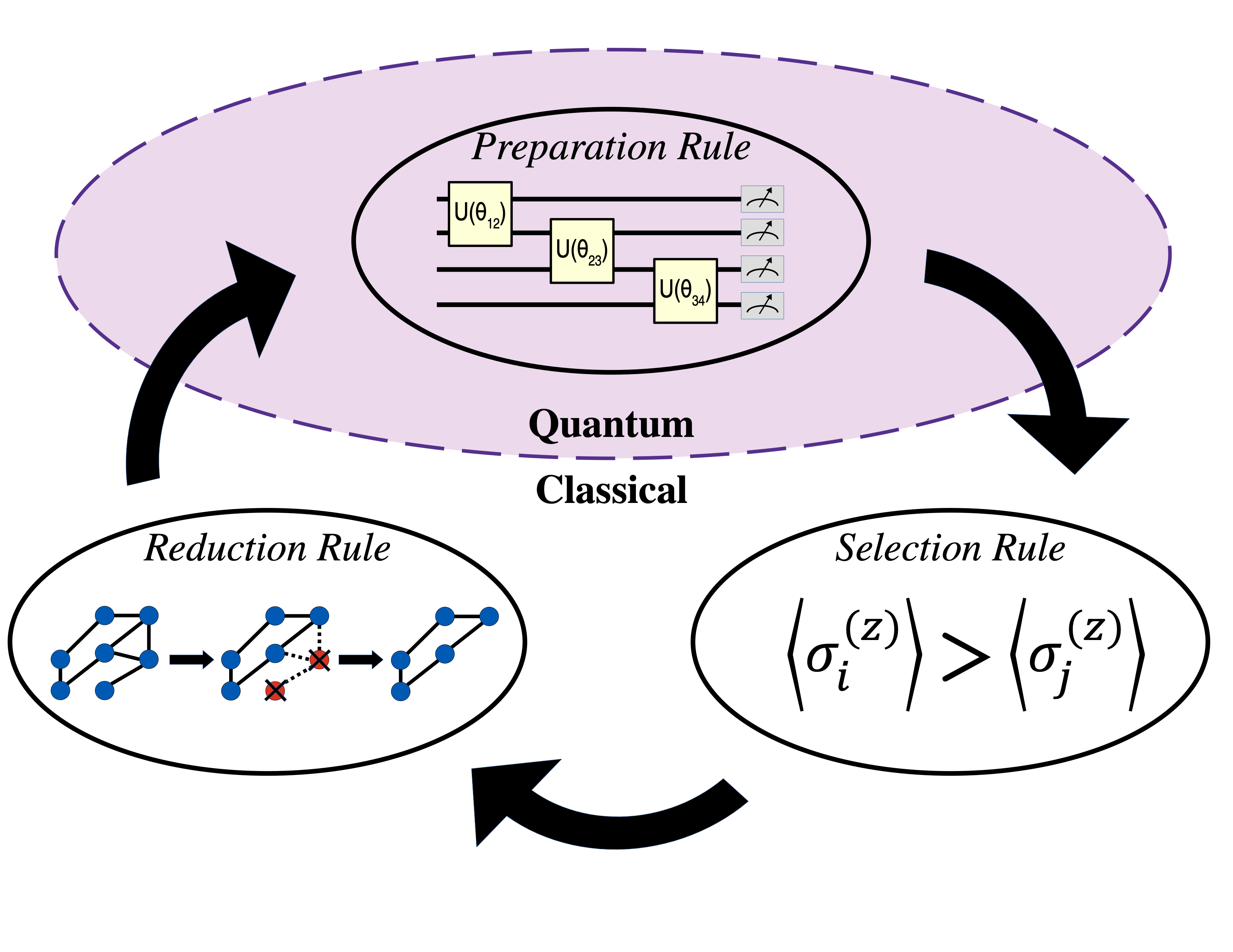}
\caption{A schematic diagram of the general structure of an Iterative Quantum Algorithm.}
\label{fig:scheme}
\end{figure}

Further work in iterative quantum algorithms has fixated not just on this circuit depth issue but also on the power inherent in the iterative loop itself, in particular the close connection to classical greedy algorithms~\cite{dupont2023quantum,Brady2023}.  
Furthermore, the nature of the iterative loops means that the quantum subroutine can be slotted out for any quantum (or often classical) optimization algorithm.  This gives us the ability to do analysis directly on the iterative loop in some cases, for instance showing that depth-1 QAOA in an iterative quantum algorithm for Maximum Independent Set has the same behavior as a standard classical greedy algorithm~\cite{Brady2023}, with improved performance observed with increasing depth.

Notably, Iterative Quantum Algorithms can be extended to constrained optimization problems, with the iterative process designed to respect the problem's hard constraints. 
First, at any step, the choice of possible problem reductions is restricted to those that do not violate any hard constraints. 
Second, after a reduction is selected, \emph{logical inference} is applied to the current subproblem and set of hard constraints to determine any further implied problem reductions, which can be significant. For example, for Maximum Independent Set, fixing a vertex to be in the set implies all neighboring vertices can also be removed from the problem graph~\cite{Brady2023}. In this work, we show how this idea applies to general classes of constrained optimization problems.

We further develop the theory of iterative quantum algorithms, explicitly focusing on both how they can handle general constraints and on how they can incorporate other elements of classical algorithms.  Previous iterative quantum algorithm work had looked almost exclusively at binary optimization problems where the natural pieces of information to fix are single variables or the relationship between two variables.  With higher locality problems, a single fixed piece of information might involve more variables and establish a more complicated relationship.  We develop methods for dealing with these more complicated relationships while simultaneously developing a framework that can handle arbitrary constraints in the problems (although, targeted algorithms for specific constraints are likely to be more efficient).

We then use these techniques to develop an iterative quantum algorithm variant of the well-known Davis-Putnam-Logemann-Loveland (DPLL) algorithm \cite{Davis_1960,Davis_1962} for deciding $k$-SAT problems.
Like its classical relatives, Quantum DPLL utilizes backtracking to yield a complete search algorithm for $k$-SAT, i.e., the algorithm is guaranteed to eventually return a satisfying assignment or certify that none exists. 
To this end we make use of the correspondence between the $k$-SAT decision problem, and max-$k$-SAT optimization problem which is amenable to quantum algorithms. 
Whereas previous iterative quantum algorithms can be viewed as exploring a single root-to-leaf path, our algorithm extends these ideas to the entire (implicitly generated) search tree. 
We further prove limiting cases in which the quantum algorithms will reproduce known DPLL performance results and provide numerics for the quantum algorithms outside of these limits. 
Our Quantum DPLL algorithm serves as proof of principle and opens the door to further synergies of IQAs with the many sophisticated classical tree search algorthms and related paradigms.

Section \ref{sec:setup} provides the setup for the problems of interest and for iterative quantum algorithms in general.  We introduce a generalized form of logical inference that can be used for complicated and constrained iterative quantum algorithm problems in Section \ref{sec:elevation}.  From there, we discuss a specific application with more tailored logical inference by introducing a quantum version of DPLL in Section \ref{sec:QDPLL}.  Then in Section \ref{sec:selection} we develop different selection rules for iterative quantum algorithms in higher locality and analytically show their behavior in limiting cases. 
We then apply these analytic results to quantum DPLL in Section \ref{sec:dpll_analysis} and show that the quantum DPLL rule reduces to a classical equivalent in limiting cases.  With that insight, we develop a more robust but non-trivial quantum DPLL rule based off existing classical rules. 
All our numerics are presented in Section \ref{sec:numerics}, and we conclude in Section \ref{sec:conclusion}.

\section{Setup}
\label{sec:setup}

Many implementations of quantum computing hardware natively support two-body interactions, which has meant that a disproportionate amount of research has focused on QAOA and Quantum Annealing with 2-local cost Hamiltonians, often without hard constraints. While methods exist for dealing with constraints, for example penalty terms, they come with tradeoffs and challenges such as increased resource overhead. As a result QAOA and Quantum Annealing are often associated with Quadratic Unconstrained Binary Optimization (QUBO) problems despite the theoretical foundations of those algorithms supporting more complicated problems.  In this section, we will outline what iterative quantum algorithms are and describe the types of problems we will focus on that relaxes both the binary and unconstrained conditions, focusing on general $k$-local clauses and interactions.

Specifically, we will consider problem Hamiltonians which can describe generic binary optimization problems~\cite{lucas2014ising,hadfield2021representation}.  In its most basic form such a Hamiltonian consists of $m$ terms, each of which has a weight $c_\alpha$ and a locality $k_\alpha$ consisting of bits contained in the set $Q_\alpha$.  These terms are simple products of the Pauli-$Z$ operators so that
\begin{equation}
    \label{eq:prob_ham_klocal}
    \hat{H}^{(T)} = \sum_{\alpha=1}^m c_\alpha \bigotimes_{j\in Q_\alpha} \sigma_j^{(z)}.
\end{equation}
Classically, in its most general form this Hamiltonian corresponds to a max-$k$-XORSAT problem instance, with each term being a weighted XOR clause that we want to satisfy.  
We refer to the products of Paulis in Eq.~(\ref{eq:prob_ham_klocal}) (i.e. a single XOR clause) as base terms that then have weights $c_\alpha$.
Generally, for any binary optimization problem other than max-$k$-XORSAT, we will have structures in the logical problem that correspond to linear combinations of these base terms.  We refer to these groupings of base terms as ``logical terms,'' and many of our techniques will deal with these logical terms directly in the selection and reduction stages of iterative quantum algorithms.  
These logical terms can have relative weights between base terms as well as a global weight that defines it strength in the full Hamiltonian.
Indeed while one could in principle design iterative quantum algorithms using the base terms directly, we demonstrate advantages of utilizing logical terms, in particular the close connection to classical algorithms.

We distinguish between the evaluated value of each base or logical term (usually restricted to be $\pm1$ or $0/1$), and the weight associated with that term.  Throughout, this paper, we take the convention that a 
term is \textit{satisfied} if and only if it evaluates to a strictly positive number.  
Since we are reformulating maximizing satisfiability problems into finding ground states of Hamiltonians (inherently a minimization problem), this means that all of the clauses that we want to be satisfied will come in with negative weights applied to the terms.
In keeping with conventions of quantum computing, we will primarily consider $\pm1$ valued variables, corresponding to the eigenvalues of the Pauli matrices.  We formulate our quantum algorithms as ground state solvers/ minimization problems, so in situations where we reward the algorithm for satisfying terms, we will include negative weights.

\subsection{Iterative Quantum optimization}
Iterative quantum algorithms consist of three 
high-level steps (see Fig.~\ref{fig:scheme} for a schematic diagram of these steps) characterized by
\begin{enumerate}
    \item Preparation Step:
    prepare a quantum state based off some ansatz (e.g. QAOA or Quantum Annealing) that is repeatedly prepared and measured to yield samples or expectation values of variables from good approximate solutions. 
    This may include an initial parameter training phase.
    \item Selection Step: 
    classically process results of preparation step to select a piece of information highlighted by the quantum algorithm, and feed it into the next step.
    \item Reduction Step: 
    constrain the problem's solution space, reducing the size of this space, possibly through variable elimination, using the information from the selection step followed by logical inference using all problem constraints.
\end{enumerate}

The selection and reduction rules may be chosen independently, allowing for greater freedom in the design of the algorithm and a larger range of possible approaches.
Furthermore rules derived from or inspired by successful classical algorithms and heuristics can sometimes be directly leveraged, as we later demonstrate. 
In particular, directly incorporating logical inference magnifies the potential reduction at each step

For selections and reductions that solely involve single qubit expectation values or two qubit correlators, as mostly considered in prior art, 
the reduction rule directly
reduces the number of variables in the problem, either by fixing a variable value or by fixing the relationship between two variables. 
In the reduction step above, we highlight the role of logical inference more broadly because more advanced forms of logical inference can facilitate extensions of iterative quantum algorithms to constrained and higher-locality problems. 
If constraints are involved in these problems, the reduction rule can respect those constraints, potentially reducing the problem further by eliminating other variables that have their values logically fixed by the selection and constraints.  
For example, for Maximum Independent Set, graph nodes are never added to the set if they conflict with any previously added nodes~\cite{Brady2023}. 
When higher locality correlators are involved in the selection (which would be natural with higher locality problems), elimination becomes harder.  Namely, a relationship between three or more bits is not enough information to eliminate one of them without potentially increasing the locality of other terms in the Hamiltonian (as was suggested in the arXiv version of the original RQAOA paper \cite{Bravyi2020}).  Because increasing the locality of other terms could lead to complications, requiring greater circuit depth and potentially ancilla qubits to encode on physical hardware, we propose an alternative approach to accounting for and using this selected and reduced information.

\subsection{Quantum subroutines}

For numerics and theory in this paper, we will mostly consider QAOA \cite{Farhi2014,hadfield2019quantum} for the preparation rule, but we emphasize that almost any quantum ansatz or suitable algorithm can be slotted into the preparation rule.  In fact, classical sampling algorithms, such as Monte Carlo techniques can also be slotted in as the selection rule with no other modifications to the algorithm structure.  
Similarly, quantum annealers and other technologies outside the quantum circuit model are also applicable.
QAOA is well-studied and has a fairly easy to implement and analyze structure, making it a good test case.
Furthermore it comes with that desirable property that with sufficiently many layers it can in principle converge to the ground state of the cost Hamiltonian.

QAOA is an optimization algorithm, structured as a variational bang-bang control problem.  The goal is to find the minimum of some classical cost function $C(x)$ over bit strings $x$.  The algorithm consists of repeatedly applying a cost Hamiltonian, whose ground state encodes the solution to the optimization problem, and a mixer Hamiltonian that is simple to construct and apply.  Usually, the cost Hamiltonian is taken to be diagonal in the computational basis (consisting of $n$ bit strings) with the cost function along the diagonal
\begin{equation}
    \hat{C} = \sum_{x\in\{0,1\}^n} C(x)\ket{x}\bra{x}.
\end{equation}
Any cost Hamiltonian of this form is uniquely expressed in the Pauli $\sigma^{(z)}$ basis as in Eq.~\eqref{eq:prob_ham_klocal}~\cite{hadfield2021representation}.
In the commonly used ground state formulation, constraints would be encoded into a constraint function $f(x)$ that evaluates to zero on the strings that satisfy the constraints and evaluates to some positive values for other strings.  The cost function can then be augmented with this penalty function via $C(x) \to C(x)+\lambda f(x)$, where $\lambda$ is some large number that can be tuned to enforce the constraints. 

The mixer Hamiltonian is usually taken to be a simple transverse field on the qubits, but this can be modified, especially to respect certain symmetries and constraints in the system \cite{hadfield2019quantum}.  Our numerics and analytics will take the mixer Hamiltonian to be
\begin{equation}
    \hat{B} = -\sum_{i=1}^n \sigma_i^{(x)},
\end{equation}
where $\sigma_i^{(x)}$ is the Pauli-X operator acting on the $i$th qubit, and the sign of $\hat{B}$ is chosen for minimization.

QAOA starts with an initial state, usually (and in our case) taken to be the ground state of the mixer Hamiltonian, $\ket{\varphi_0}$ and applies $p$ layers of quantum circuit.  In each layer, we apply $\hat{C}$ for a fixed time and then $\hat{B}$ for a fixed time.  These times are parameterized by angles $\vec{\gamma}$ and $\vec{\beta}$ respectively so that the resulting state is
\begin{equation}
    \ket{\psi(\vec{\beta},\vec{\gamma})} = e^{-i\beta_p\hat{B}}e^{-i\gamma_p \hat{C}}
    \dots
    e^{-i\beta_1\hat{B}}e^{-i\gamma_1 \hat{C}}
    \ket{\varphi_0}.
\end{equation}.

Then the cost function is evaluated on this state 
\begin{equation}
    J(\vec{\beta},\vec{\gamma}) = \bra{\psi(\vec{\beta},\vec{\gamma})}\hat{C}\ket{\psi(\vec{\beta},\vec{\gamma})}.
\end{equation}
Based off the evaluated $J(\vec{\beta},\vec{\gamma})$, a classical outer loop algorithm then updates the angles, $\vec{\beta}$ and $\vec{\gamma}$, trying to minimize the value of $J(\vec{\beta},\vec{\gamma})$.  This hybrid back and forth between the quantum computer and the classical outer loop continues until the cost function stabalizes to a minimum that hopefully corresponds to the ground state energy of $\hat{C}$.

\subsection{Classical Optimization Problems}

We consider two general classes of optimization problems for our numerics and analytics in this paper, but much of the theory developed here can be extended to other problems.

The first problem we consider is the one where each logical term is simply a product of Pauli-$Z$s evaluated on qubits in a hyperedge so that the logical terms are identical to base terms.
In computer science, this problem is equivalent to max-$k$-XORSAT where we have a collection of clauses, that each consist of multiple variables or negations of variables that are all XORed together.
In max-$k$-XORSAT, each of these clauses has at most $k$ variables, and the goal is to find assignments for the variables that maximize the number of satisfied clauses.
For $k=2$, this problem is the MAXCUT problem.
We further consider a weighted version of this where each clause gives a specific reward for being satisfied, and the goal is to maximize the rewards.  We choose our weights randomly in the range $[-1,1]$, and we also select the edges in the hypergraph randomly, choosing $m$, the number of hyperedges, at the beginning.

The second problem takes more advantage of the logical terms needed for our problem by considering max-$k$-SAT.  This is a classical problem from computer science where we have a set of $m$ clauses, each containing at most $k$ variables ORed together (we consider the subvariant where each clause contains exactly $k$ variables).  Each variable is either a bit value or the negation of that bit value (referred to as the variable having positive or negative polarity), and the goal of the algorithm is to find the maximum number of clauses that can be satisfied simultaneously by a given bit assignment.  Furthermore, the clauses can be weighted, and we choose weights in the range $[0,1]$.

Max-$k$-SAT is readily formulated in terms of $0/1$ binary variables, but we need to convert this to $\pm1$ variables for the quantum formulation.  Usually to represent one of these clauses numerically, you can take each variable (including positive and negative polarities) in the OR clauses and construct a mathematical equation.  You start by adding the values of all the variables together, then subtracting all the pairwise multiplications of variables, then adding the three-wise multiplication of variables and so on until you reach terms including all the variables.  To convert these from $x\in\{0,1\}$ to $z\in\{-1,1\}$ variables, we then need to use $x = \frac{1}{2}(z+1)$ which takes $1\to1$ and $0\to-1$ (note that this convention is opposite that used in some branches of quantum computing).  Further $\bar{x} = \frac{1}{2}(-z+1)$ for negations.  This allows us to reformulate this problem entirely in terms that we can implement in quantum Hamiltonians.  Also note that since we prefer to do a minimization problem, we will bring in every clause with a negative weight.

\section{Hard Constraints and Logical Inference}
\label{sec:elevation}

The Reduction Step is generally a reduction in the solution space, and each reduction does not necessarily imply variable elimination.  When we fix information in the problem, we are now enforcing that information to hold for any solution we find.  But this description is just that of a problem constraint, so when the reduction rule is not specific enough to lead to full variable elimination to satisfy it, we elevate the information in the reduction rule to consider it alongside the other constraints in the problem. Hence reduction steps may be viewed generally as the introduction of additional problem hard constraints, followed by any inferred variable eliminations.

To give an example, consider the problem Hamiltonian in Eq.~(\ref{eq:prob_ham_klocal}), which as explained directly corresponds to a classical cost function and which has its logical terms equal to its base terms.  

Here we consider the case of using the logical terms of the Hamiltonian itself for the selection rule.  We denote the logical terms in the Hamiltonian as $h_\alpha$, such that $\hat{H}^{(T)}= \sum_\alpha w_\alpha h_\alpha$, possibly with weights $w_\alpha$.
While the selection rule can consider any set of features to select on, a natural option is to choose features present in the Hamiltonian or the logical problem. 
As an example we consider selecting from the set of $h_\alpha$, picking the logical term with the largest absolute expectation value.  This can then chain into a reduction rule that fixes that term in the Hamiltonian to be equal to the assignment implied by its expectation value (here assuming that the term can take $\pm1$ values, not $0/1$ values)
\begin{align}
    \max_{\alpha} \left|\avg{h_\alpha}\right| \; \Rightarrow
    \; h_{\alpha^*}=\text{sgn} \langle h_\alpha \rangle 
\end{align}
where $\alpha^*$ refers to the maximizing index.
Classically, this corresponds to fixing the value of a clause.
The resulting rule is not enough information to fix any of the bits unless the number of bits in the term, $k_{\alpha^*} = |Q_{\alpha^*}| \leq 2$, so instead we remove this term from the Hamiltonian and add it to the constraint, so that a newly added constraint would have the form 
\begin{equation}
    \hat{H}^{(C)}_i  = \text{sgn}(\avg{h_{\alpha^*}}) \; h_{\alpha^*}.
\end{equation}
where $i$ indexes over the set of all constraints in the problem, $\mathcal{C}$, either from the original formulation or from this elevation style.
The final constraint Hamiltonian would be a linear combination of all the individual constraint Hamiltonians.  If a penalty formulation is being used, then the implemented Hamiltonian for the next round of IQO would then be
\begin{equation}
    \hat{C} = \hat{H}^{(T)} + \lambda\sum_{i\in\mathcal{C}}\hat{H}^{(C)}_i,
\end{equation}
where $\lambda$ is a suitably large multiplier.

But elevating a reduced term to a constraint is not enough because we need to know how that newly added constraint interacts with the rest of the constraints, either from previous iterations or from the natural implementation of the problem.  It is possible that the newly added constraint will provide enough information in conjunction with previous constraints that we are able to eliminate variables or other terms in the Hamiltonian.  This is where logical inference comes in.

There are plenty of general purpose logical inference algorithms for binary variables that can take on $0/1$ values, most of which operate by finding patterns in the truth tables for satisfying strings, such as the Quine-McCluskey Algorithm \cite{Quine01101952,Quine01111955,McCluskey1956}.  In Appendix \ref{app:logical-inference}, we use this same truth table pattern finding strategy to develop algorithms for our general setting. In practical applications the inferencing rules or algorithm employed may be problem structure dependent, and often this additional information can drastically speed up the process of logical inference.

For problems that start out with no inherent constraints, the constraints will arise just from elevated terms, one at a time with reductions made as we go.  The constraint Hamiltonian will itself have structure, and for hypergraph problems, like we consider in this paper, the constraint Hamiltonian will itself have a corresponding hypergraph structure.   While there are cases where a variable elimination is not possible for a substantial amount of time, especially for higher locality problems, in many cases, the hypergraph constructed by these constraints will remain sparse and largely disconnected.  We only need to worry about finding satisfying strings in each of these disconnected hypergraphs, and all evaluations of logical inference can similarly be restricted to each disconnected sub-hypergraph.

A precise description of our full logical inference routines can be found in Appendix \ref{app:logical-inference}.  This algorithm relies on truth tables to identify variables that can be eliminated due to the logical inference rules. At each reduction step, the logical inference algorithm  considers $n'$ variables represented in the constraints with $m'$ strings that satisfy those constraints with a worst case runtime of $\mathcal{O}(n'^2 m')$ and a best case runtime of $\Omega(n'm')$.  Again, for problems with more structure in their constraints, the logical inference algorithm can be tailored, resulting in shorter runtimes.

\section{Quantum DPLL}
\label{sec:QDPLL}

Iterative Quantum Algorithms, like related classical greedy algorithms, are performance limited by the quality of choice they make at each iteration. Generally, for any choice of selection and reduction rule, we do not expect either approach to exactly solve NP-hard problems efficiently. Classically, this has inspired a wide variety of solution-tree search algorithms, which may require more than polynomially scaling time, but are guaranteed to eventually find and certify the optimal solution. We next show that the basic ideas of Iterative Quantum Algorithms may be directly integrated into the powerful classical algorithmic paradigms. In particular our approach facilitates the effective use of fixed quantum resources within a hybrid classical framework that may call the quantum device a superpolynomial number of times.

The Davis-Putnam-Logemann-Loveland (DPLL) algorithm \cite{Davis_1960,Davis_1962} is a well-known and highly successful family of classical algorithms for determining satisfiability of $k$-SAT decision problems.  Even state-of-the-art modern SAT solvers use algorithms derived from DPLL (such as CDCL \cite{569607,769433,10.5555/1867406.1867438} used in several successful recent solvers in annual SAT competitions \cite{SAT_Competition,BiereFallerFazekasFleuryFroleyks-CAV24}).  This algorithm uses single variable fixing along with problem-specific logical inference to reduce the problem at each step. This means that its structure is amenable to the form of iterative quantum algorithms that we are employing, combined with \textit{backtracking} to enable complete coverage of the search tree so as to guarantee that upon termination the correct decision is returned. 

Here we combine and generalize the DPLL algorithm with iterative quantum algorithms. In the setting of solution tree search, the selection and reduction steps become \textit{branching rules}, which determine the structure of the search tree and the order in which nodes are explored. 
In the previous section, we considered using logical terms in the Hamiltonian for logical inference which works well when we are trying to maximize the number of satisfied terms.  However, in satisfiability (SAT) decision problems, the goal is to have all terms satisfied, meaning that such a term fixing scheme would just be  a less useful matter of order.  Instead, DPLL focuses on selecting and reducing variables.

At each step a variable value is fixed and inference applied, resulting in a smaller SAT decision, for which the process is repeated. When a branch is ruled out, backtracking allows the algorithm to recursively explore the variable value selections not taken. 
Here we focus on the SAT decision problem in order to elucidate the main ideas and results. Moreover decision properties have desirable properties for direct comparison to other algorithms such as classical DPLL; in particular, that the output is either correct or not, and so direct runtime comparisons are meaningful. 
Nevertheless similar ideas can be directly extended to MaxSat as well as other optimization problems; we explore this case in detail in companion work.

The SAT problem we are trying to solve is usually written in conjunctive normal form where we are trying to evaluate the truth of a statement consisting of ANDs of clauses, each of which consists of OR'd variables (or their negations).  Here we will use $x_i$ to represent $0/1$ variables with a bar over a variable denoting its negation.  The basic structure of DPLL consists of the following: 
\begin{enumerate}
    \item Choose a variable to branch on and select the initial value (0 or 1) of that variable to choose.
    \item Reduce the problem by fixing the chosen variable and value
    \item Iteratively reduce other variables that have a necessary value by applying logical inference.
    \item If all variables have been reduced with no contradictions, return True for the algorithm.
    \item If there are still variables that cannot be reduced and there are no contradictions, calculate DPLL applied to the new smaller problem.
    \item If a contradiction was reached or if DPLL applied to the smaller problem in step (5) returned False, return the problem to its state before step (2) and repeat steps (2)-(5) with the opposite value for the chosen variable.
    \item If both values have been tried with no satisfying assignment found, return False.
\end{enumerate}

Classically, some branching rule or heuristic for selecting both the variable and value must be specified for Step 1, with the resulting performance highly dependent on this choice. Importantly, this choice determines the order of variables and values explored, and thus determines the structure of the solution search tree.
The logical inference of Step 3 utilizes two standard rules for SAT, and hence avoids the overhead of the fully general inference procedure we detail in Appendix~\ref{app:logical-inference}. 
The resulting reductions rules are \textit{unit propagation} and \textit{pure literal elimination}.  As every clause must be satisfied in a SAT solution, if a clause contains only a single variable, we must fix that variable to satisfy the clause performing such fixes immediately is called unit propagation. Similarly, pure literal elimination checks if a variable only occurs with a single polarity in every clause, and if so that variable can be fixed correspondingly.
These two reduction are iteratively implement at a given step until they no longer apply, as often they will combine and cascade to yield further reductions.

\subsection{Quantum-enhanced Branching Rule}

For DPLL, the choice of branching rule (Step 1) can make a major difference in the performance of the algorithm \cite{Hooker_1991,Hooker1995}.  Heuristically, it is beneficial to branch based off variables that will lead to a large number of satisfied clauses, and this heuristic inspired some of the early classical branching rules.  However, the structure of this branching rule is functionally identical to the preparation and selection rules we are using in iterative quantum algorithms.  Therefore, we can easily slot in a quantum subroutine for this branching rule.

To see what such a quantum branching rule would look like, we can consider selecting a single Pauli-Z to branch with.  A problem is that it is not easy to formulate a satisfiability decision problem in a form easily ingestible by a quantum computer~\cite{hadfield2021representation}.  On the other hand, we can easily formulate the max-$k$-SAT problem whose ground state energy directly determines the satisfiability of the corresponding $k$-SAT problem.

In the max-$k$-SAT problem, we are again given a set of disjunctive clauses, but now we seek to decide if there exists an assignment satisfying at least $\ell$ clauses. In the corresponding optimization problem, we seek a string satisfying as many clauses as possible. Each clause can be mapped to a Hamiltonian as in Eq.~\ref{eq:prob_ham_klocal} using standard techniques and the substitution $x_i\to (1+\sigma_i^{(z)})/2$. For instance, a clause consisting of $k$ variables drawn from a subset of variables $i\in X$ (here as an example all with positive polarity), can be written numerically as 
\begin{equation}
    \label{eq:SAT_clause}
    \frac{1}{2^k}\left(2^k-1+\sum_{m=1}^k(-1)^{m-1}\sum_{\substack{Y\subseteq X\\|Y|=m}}\left[\prod_{i\in Y} \sigma_i^{(z)}\right]\right).
\end{equation}
If the $j$th variable has a negative polarity, we can just replace $\sigma_j^{(z)}\to-\sigma_j^{(z)}$ everywhere in the expression to account for this.
The resulting Hamiltonian for max-$k$-SAT can then be used in a quantum or classical subroutine to produce a ranking of the bits in the problem.
Later in Section \ref{sec:dpll_analysis}, we apply analytic tools that we develop in the next section to determine connections between this quantum branching rule and well known classical branching rules.

Note that a quantum enhancement to DPLL has been proposed in the past \cite{Montanaro2016}.  There the proposal was to enhance the backtracking in DPLL with quantum algorithms within a fully quantum-coherent tree-search setting, i.e., a deep circuit on a fault-tolerant quantum device; whereas here we are focused on just the branching rule, which utilizes a quantum subroutine embedded within a classical tree search.

\section{Selection Rules at Low Depth}
\label{sec:selection}

Before we proceed to analyze Quantum DPLL and other methods that can be improved with iterative quantum algorithms for higher locality problems, it is useful to establish analytic tools for determining limiting cases of these algorithms.  This section primarily focuses on results for $p=1$ QAOA, especially in limiting cases, where analytical results can be obtained.  The main body of this section focuses on QAOA results for single $Z$ expectation values, but the appendices focus on how to extend these results to higher order correlators. While deeper QAOA circuits as well as other sophisticated algorithms are desirable in practice, they are both challenging to analyze as well as more daunting for near-term implementations.

With any iterative quantum algorithm, the choice of selection rule is highly problem dependent, and will greatly influence how the problem is solved.
As explained, a choice that is always natural and relatively easy to analyze, is just looking at the expectation values of single bits (qubits) in the sampled distribution and ranking them based on their magnitudes. This choice is easy to implement and universal to all problem types, up to respecting the current problem constraints.

\subsection{Single-$Z$ QAOA $p=1$ Expectation Values}

Here we show analysis of single-qubit expectation values for QAOA protocols. 
We consider the more general case of correlators of multiple variables in Appendix \ref{app:higher-locality}.

For the case of using $p=1$ QAOA as the preparation rule, we can carry out a path-sum analysis, analogous to the one performed in Ref.~\cite{Brady2023}.  Consider a depth $p=1$ QAOA algorithm where we are looking at expectation values:
\begin{equation}
    J_j(\vec{\beta},\vec{\gamma}) = \bra{\varphi_0}\hat{U}^\dagger(\vec{\beta},\vec{\gamma})\sigma_j^{(z)}\hat{U}(\vec{\beta},\vec{\gamma})\ket{\varphi_0},
\end{equation}
where $\hat{U}(\vec{\beta},\vec{\gamma}) = \prod_{i=1}^p e^{-i\beta_i\hat{B}}e^{-i\gamma_i \hat{C}}$.  The Hamiltonians are $\hat{B}$, a simple mixer, taken here to be a transverse magnetic field, and $\hat{C}$, the problem Hamiltonian that encodes the classical cost function of interest along its diagonal.  We will refer to the classical cost function as $C(z)$ evaluated on a bit string $z$.  Ref.~\cite{Brady2023} worked this out for general $C(z)$ in their Eq.~(A6):
\begin{align}
    J_j(\beta,\gamma) &= \frac{1}{2^{n+1}}\sum_{z\in\{-1,1\}^n} 
    \sum_{z'_j=\pm z_j}
    e^{i\gamma (C(z')-C(z))}\\\nonumber
    &
    \times\left(e^{2i\beta}(-1)^{\frac{1+z_j}{2}} + 
            e^{-2i\beta}(-1)^{\frac{1+z'_j}{2}}\right).
\end{align}
where $z$ and $z'$ are the same other than the $j$th bit, with a sum going over whether that bit is the same or different between them.

If $z'_j = z_j$, then $C(z')-C(z)=0$, but if $z'_j = -z_j$, we end up picking out all the physical terms in the Hamiltonian that contain exactly one copy of $z'_j$.  Note that for any of the $z_k$, $z_k^2 = 1$, so we don't have to worry about terms containing more than one copy of $z_j$.  We will refer to all the physical terms that contain one copy of $z_j$ as $z_j C'_j(z)$, and we will further separate out $C'_j(z) = c_j + C_j(z)$ into terms that are constant and terms that are dependent on other variables.  With these definitions in mind, the case where $z'_j = -z_j$ gives us $C(z')-C(z)=-2 z_j\left(c_j+C_j(z)\right)$.  Therefore, the expression can be simplified to 

\begin{align}
    J_j(\beta,\gamma) =& \frac{1}{2^{n}}\sum_{z\in\{-1,1\}^n} 
    \bigg(
    (-1)^{\frac{1+z_j}{2}}\cos(2\beta)\\\nonumber
    &
    +i (-1)^{\frac{1+z_j}{2}} e^{-2i\gamma z_j\left(c_j+C_j(z)\right)}\sin(2\beta)\bigg).
\end{align}

We refer to the neighborhood of nodes that share a hyper edge with 
node $j$ as $N(j)$ and refer to the size of this set as $d_j = |N(j)|$.  Doing the sum over all nodes not in $N(j)\bigcup\{j\}$ is trivial and just gets us a factor of $2^{n-d_j-1}$.  Our expression explicitly calls out all instances of $z_j$ itself, so we can do the sum over $z_j=\pm1$ explicitly as well
\begin{align}
    J_j(\beta,\gamma) = \frac{-1}{2^{d_j}}\sin(2\beta)\sum_{\substack{z_k=\pm 1\\k\in N(j)}} \sin\left(2\gamma\left(c_j+C_j(z)\right)\right).
\end{align}

Since $c_j$ does not depend on the variables, we can use the angle addition formulas to separate it out
\begin{align}
    \label{eq:J_p=1}
    J_j(\beta,\gamma) = &\frac{-1}{2^{d_j}}\sin(2\beta)
    \bigg(
    \sin(2\gamma c_j)\sum_{\substack{z_k=\pm 1\\k\in N(j)}} \cos\left(2\gamma C_j(z)\right)\nonumber\\
    &
    +
    \cos(2\gamma c_j)\sum_{\substack{z_k=\pm 1\\k\in N(j)}} \sin\left(2\gamma C_j(z)\right)
    \bigg).
\end{align}

$C_j(z)$ contains no constant terms and only combinations of binary variables, so in general we can say that $\sum_{\substack{z_k=\pm 1\\k\in N(j)}} C_j(z) = 0$, but this property does not hold in general after we have applied sines and cosines to them.  The expression we have lends itself to some simplifications, for instance if $c_j=0$, but without further problem specific information, this cannot be simplified further.

However, the expression in Eq.~\ref{eq:J_p=1} does lend itself well to either small $\gamma$ approximations or direct calculation.  Exactly calculating this expression can require $\mathcal{O}(2^{d_j})$ time in the worst case.  We would want to calculate this quantity for every one of the $n$ nodes in the hyper-graph, but if the maximum degree of a node is $d = \max_j d_j$, then exactly calculating the results of a QAOA-1 preparation rule with single-$z$ expectation values just requires $\mathcal{O}(n 2^d)$ time.  Often the degree of nodes in the hyper-graph $d\ll n$, so for sparser hyper-graphs, this could be a quite doable calculation.

In Appendix \ref{app:2-local}, we show how this general result can reduce to known results from the literature for 2-local problems.

\subsubsection{Small $\gamma$ Approximation}

Furthermore, we can consider the case when $\gamma$ is a small quantity and simplify Eq.~\ref{eq:J_p=1}.  A small $\gamma$ approximation is common in theoretical QAOA analysis \cite{hadfield2022analytical,Wurtz_2022}, but is not quite justified in most practical situations \cite{Zhou_2020,Pagano_2020}.  Here we are performing this analysis not to see exactly what QAOA will do but to get a sense of the quantities it cares about and to potentially back out classical analogous rules.

Starting from Eq.~\ref{eq:J_p=1}, let's consider a third order expansion in $\gamma$:

\begin{align}
    J_j(\beta,\gamma) = &\frac{-1}{2^{d_j}}\sin(2\beta)
    \sum_{\substack{z_k=\pm 1\\k\in N(j)}}\bigg(
    2\gamma (c_j + C_j(z))\\\nonumber
    &
    -\frac{4}{3}\gamma^3 (c_j+C_j(z))^3
    \bigg)+\mathcal{O}(\gamma^5).
\end{align}

But this expression can be simplified by noting that the sums over bit strings at the first order expansion can separate linearly over terms.  Furthermore, because each of these terms is just a product of $\pm1$ variables being summed over, these sums will all necessarily evaluate to zero.  Thus, these terms only come in starting at third order in $\gamma$:
\begin{align}
    \label{eq:small_gamma}
    J_j(\beta,\gamma) &= -2\gamma\sin(2\beta) c_j \\\nonumber
    &
    +\frac{4\gamma^3\sin(2\beta)}{3\cdot2^{d_j}} \sum_{\substack{z_k=\pm 1\\k\in N(j)}}(c_j+C_j(z))^3
    +\mathcal{O}(\gamma^5).
\end{align}

So to first order in $\gamma$, depth one QAOA just cares about the linear terms for selection rules.  In max-XORSAT, this linear term is zero, for Maximum Independent Set, this term is proportional to the degree of the selected node, and for max-SAT, this is related to the difference in the number of terms containing a variable versus its negation.



\section{Quantum DPLL vs. Classical DPLL}
\label{sec:dpll_analysis}

In this section, we combine together the IQA formulation of Quantum DPLL, developed in Section \ref{sec:QDPLL}, with the analytic tools developed in Section \ref{sec:selection} to determine classical branching rules that correspond to our quantum versions with $p=1$ QAOA.  Further, we take well known classical branching rules and develop quantum branching rules to emulate them in the low $p$ limit. While deeper QAOA circuits are desirable in practice, they quickly become intractable to analyze using most known methods or outside of special cases.

Indeed, while a $p=1$ QAOA is the simplest possible case, it can be illustrative when trying to determine what an iterative quantum algorithm would do in this case.  Simply plugging the resulting Hamiltonian into Eq.~(\ref{eq:J_p=1}) is not horribly insightful itself, but we can take the further approximation to a small $\gamma$ limit to use Eq.~(\ref{eq:small_gamma}).  As a cautionary note, the small $\gamma$ limit is not observed in practice for optimized QAOA.  While usually $\gamma<1$, we do not have $\gamma\ll1$, so the limit we are introducing here is not exactly representative of how QAOA will actually behave.  Still, this limit gives some indication of the things that QAOA will care about.

Looking at just the first order in $\gamma$ for $p=1$ QAOA, we can see that it only depends on single variable terms in the Hamiltonian.  From Eq.~(\ref{eq:SAT_clause}), we can see that each clause will contribute linear terms to the Hamiltonian that are weighted by $\frac{1}{2^k}$ if the clause has locality $k$.  Therefore, if we refer to the clauses in the Hamiltonian by $C_{\alpha}$, with $\alpha$ indexing over all clauses, and with each clause having locality $k_\alpha$, then we get the linear weight on the $j$th qubit coming out to be
\begin{equation}
    \label{eq:firstorder}
    c_j = \sum_{\substack{C_\alpha\\ j\in C_\alpha}} 2^{-k_\alpha}-
          \sum_{\substack{C_\alpha\\ \bar{j}\in C_\alpha}} 2^{-k_\alpha},
\end{equation}
where the first sum goes over all clauses that contain the bit $j$ with positive polarity and the second sum going over all clauses that contain the bit $j$ with negative polarity.

The ranking of bits to branch off of would then be linear in this $c_j$ quantity up to first order in $\gamma$.  But this quantity, $c_j$, should look familiar to people versed in DPLL since it is already used to rank bits in a well known possible classical branching rule.  This quantity can be derived based off the first order probability expansion based off the satisfaction hypothesis that branching to maximize subproblem satisfiability is advantageous \cite{Hooker1995}.  This first-order probability rule is similar to the original Jeroslow-Wang rule (JW rule) \cite{10.1007/BF01531077} that branches off the literal (meaning the coincidence of a bit and polarity) $l$ that maximizes
\begin{equation}
    \label{eq:JW}
    c_l = \sum_{\substack{C_\alpha\\ l\in C_\alpha}} 2^{-k_\alpha}.
\end{equation}

Even though the JW rule is itself a truncation of the fuller first-order probability rule, it is known classically that the JW rule outperforms the first-order rule in most numerical settings \cite{Hooker1995}, so we would heuristically expect $p=1$ QAOA to fall somewhere in between the performance of these two.

As a note, the DPLL algorithm is not an approximate algorithm and is designed to find or disprove a satisfying assignment.  The real question of branching rule choice is how much of the search space needs to be explored to reach that conclusion.  For a satisfiable problem, an ideal case would need to explore only a single branch; whereas for an unsatisfiable problem, multiple branches need to be explored to disprove satisfiability.  The number of branches explored can vary vastly between branching rules, with a better branching rule exploring fewer branches to reach a conclusion.

In the numerics section, we will consider QAOA with $p=1$ and $p=2$ as well as both the JW rule and the first-order probability rule numerically on random $k$-SAT instances to judge performance.  In order to have a fair comparison, we will consider the number of branches explored rather than wall-clock time because the hardware (or in our case quantum simulator) used to run the quantum algorithm will be vastly different than the hardware used to run corresponding classical rules.

\subsection{Classical-Inspired Quantum Branching}
\label{ssec:classical-inspired}

Here we demonstrate how we may further tailor the problem encoding and quantum algorithm to reproduce a target algorithm result.  We focus on a known classical branching rule as our target and produce a QAOA algorithm that mimics it in particular parameter limits.

The JW rule given by Eq.~(\ref{eq:JW}) usually outperforms the first-order rule, Eq.~(\ref{eq:firstorder}), so it is natural to try to find a quantum algorithm that mimics the JW rule instead of the first-order rule.  Unfortunately, we cannot do this with our current problem formulation since the single-Z expectation values will necessarily include both polarities of the variables in their evaluation.  We can however alter the form of the problem Hamiltonian to more fully separate out the different polarities, allowing us to get a QAOA Hamiltonian that does reproduce Eq.~(\ref{eq:JW}) in the $p=1$, small $\gamma$ limit.

First, we need to make sure that the different polarities of variables in the SAT problem can be addressed individually.  To do this, we can introduce new auxiliary variables.  So for each variable in the problem, we would split it into two variables representing the different polarities of the variable so that $(+\sigma_i^{(z)})\to\sigma_i^{(z)}$ and $(-\sigma_i^{(z)})\to\sigma_{i+n}^{(z)}$.  This means that our SAT problem formulation currently only has variables with positive polarity.
To complete the transition from the original SAT problem to this new representation, we need to introduce new clauses for every split set of variables with these clauses being true if and only if those split variables are the not of each other. 

In terms of Pauli matrices, this means that we need new terms in the Hamiltonian of the form
\begin{equation}
    \label{eq:dpll_penalty}
    \sum_{i=1}^n\frac{1}{2}(-\sigma_i^{(z)}\sigma_{i+n}^{(z)}+1)
\end{equation}
with each of these terms being satisfied (giving one) when the two variables are anti-correlated and being unsatisfied (giving zero) when they are correlated.  With this addition to the Hamiltonian, we can see that our new problem is satisfiable if and only if the original problem is satisfiable.  It would be easy to eliminate these new clauses by using them to collapse the split polarity variables, but keeping them split allows us to address each of the polarities separately during QAOA measurements.  Pure literal elimination needs to be modified to keep track of the pairs of positive and negative split variables so as to not immediately eliminate this split behavior, but otherwise DPLL can be run as normal.

Now returning to the $p=1$, small $\gamma$ limit, Eq.~(\ref{eq:small_gamma}), we still want to address $c_j$, the coefficients on the linear terms for the Pauli Z matrices, as the first order contribution.  But our penalties introduced in Eq.~(\ref{eq:dpll_penalty}) do not have a linear term and so are invisible in this limit.  Therefore, with the polarities separated, we will exactly reproduce the JW rule in this limit with Eq.~(\ref{eq:JW}).

Furthermore, as we use this version of QAOA as the DPLL branching rule, the elimination of one of the polarity split variables will automatically eliminate the other half from the problem via unit propagation.  We do expect this form of DPLL to require more unit propagation steps in general, but the costly steps, especially for the quantum augmented version of the algorithm will be the branching steps.

In the numerics section, we also consider this split version of QAOA when evaluating the performance of various branching rules.

\section{Numerics}
\label{sec:numerics}

\subsection{Max Satisfiability Problems}

For the maximum satisfiability problems, we consider two algorithms for these $k$-local Hamiltonians that are initially unconstrained.
The first method uses single-$z$ expectation values in its selection rule, just picking the qubit with the highest absolute value expectation value.  The reduction rule then fixes that bit to be the sign of its expectation value.  If there were no constraints in the original problem, this can be done simply with no need for elevating things to constraints or handling logical inference.

The second method runs the selection rule based off the terms that are present in the problem.  So we are ranking the logical clauses in our problem based off which one has the highest expectation value when evaluated on samples from a QAOA produced state.  We do not consider the weights attached to the clauses, and based on the problems we are considering, the clauses evaluate either to $\pm 1$ (for max-$k$-XORSAT) or to $0/1$ (for max-$k$-SAT), independent of the size of the clauses.  This version of the algorithm considers a lot more information about the interactions in the Hamiltonian with each expectation value corresponding to a hyper-edge (or set of hyper-edges) in a graph representing the problem.  As we discussed in previous sections, this kind of selection rule benefits from a reduction rule that relies on logical inference.  For this version, we use the full procedure of elevating selected terms to constraints to keep track of them until logical inference leads to a variable reduction.
Note that this procedure is significantly slower than considering single-$Z$ expectation values both because it requires more classical computation for logical inference and because more quantum calls are needed since we are not reducing the problem size at each step.

\subsection{Max Satisfiability Results}

\begin{figure}
\begin{center}
\begin{subfigure}{0.48\textwidth}
    \includegraphics[width = 1\linewidth]{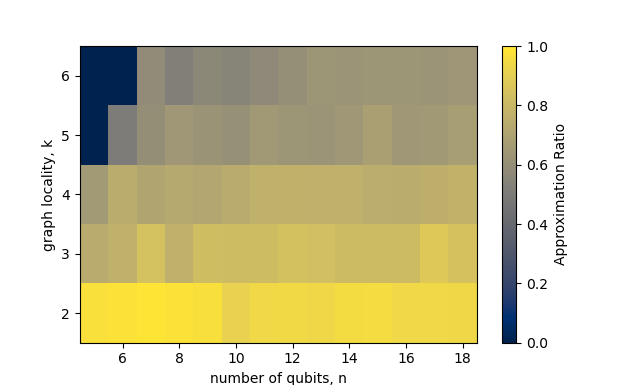}
    \caption{Single-$Z$}
\end{subfigure}
\begin{subfigure}{0.48\textwidth}
    \includegraphics[width = 1\linewidth]{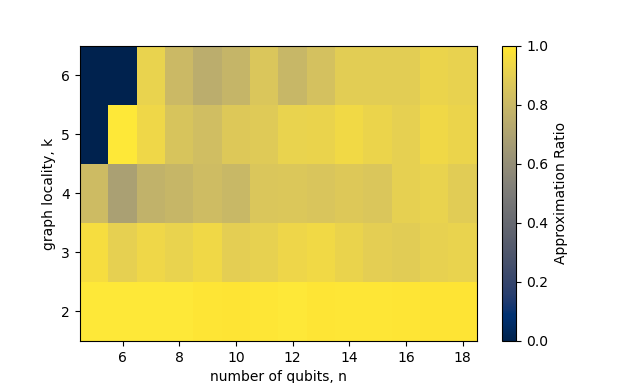}
    \caption{Logical}
\end{subfigure}
\begin{subfigure}{0.48\textwidth}
    \includegraphics[width = 1\linewidth]{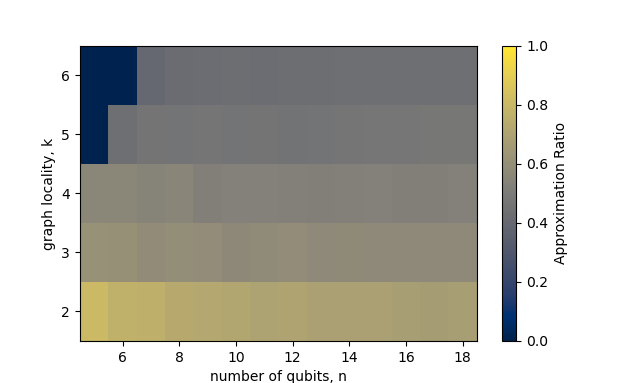}
    \caption{Base QAOA-2}
\end{subfigure}
\end{center}
\caption{The results of iterative quantum algorithms applied to max-kXORSAT using logical inference to eliminate terms at each step without increasing locality, using QAOA-$2$ as the preparation rule.  The approximation ratio is the achieved energy divided by the true ground state energy, and the graphs were created randomly with hyperedge weights in the range $[-1,1]$.  The density of hyperedges (number of hyperedges divided by number of nodes) $m/n = 1$.  The plots are for different selection and reduction rules: a) selecting and reducing single Pauli-$Z$ expectation values, b) selecting and reducing logical terms in the Hamiltonian, and c) the base results of QAOA-2. 
}
\label{fig:max-kXORSAT}
\end{figure}

In the numerics, we use $p=2$ QAOA as the preparation rule.  In Fig.~\ref{fig:max-kXORSAT}, we plot results for max-$k$-XORSAT, choosing, $m=n$ which is the density of edges that corresponds to the critical density where the phase transition from fully satisfiable to unsatisfiable occurs \cite{PITTEL_SORKIN_2016}.  The color coding corresponds to the approximation ratio averaged across 50 randomly generated instances per data point (the same instances were used for both graphs).  The upper left corner is blank due to no data points being attempted in this region.  We show results using either Logical Hamiltonian terms or single Pauli-Z terms in the selection and reduction as well as the results for base QAOA.  While both IQA results improve dramatically on base QAOA, including the full logical terms with more information about the system in the selection and reduction rules leads to better overall results.

We do not present results for max-$k$-SAT because at the system sizes considered, all our algorithms can find the maximum satisfying string with incredibly high likelihood to the point where the plots are not informative.  Such a study will need to wait for comparison with future, larger and fault-tolerant quantum hardware.

\subsection{Quantum DPLL Results}

\begin{figure}[t]
\includegraphics[width = 0.48\textwidth]{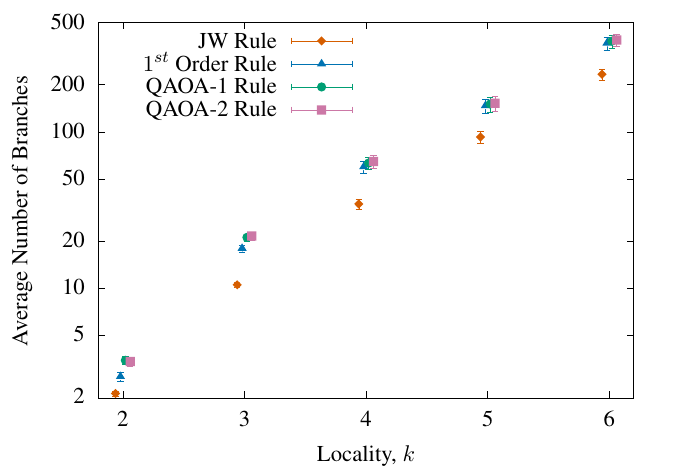}
\caption{The results of DPLL algorithms with various quantum and classical branching rules applied to random $k$-SAT problems with $n=15$ variables and $m = \lfloor n\ln(2) 2^k\rceil$ clauses.  The $y$-axis shows the average number of branching points that were needed by the DPLL algorithm to reach its conclusion.  Each point is averaged over 50 random instances, and the error bars represent the standard error of the mean.
}
\label{fig:DPLL15}
\end{figure}

\begin{figure}[t]
\includegraphics[width = 0.48\textwidth]{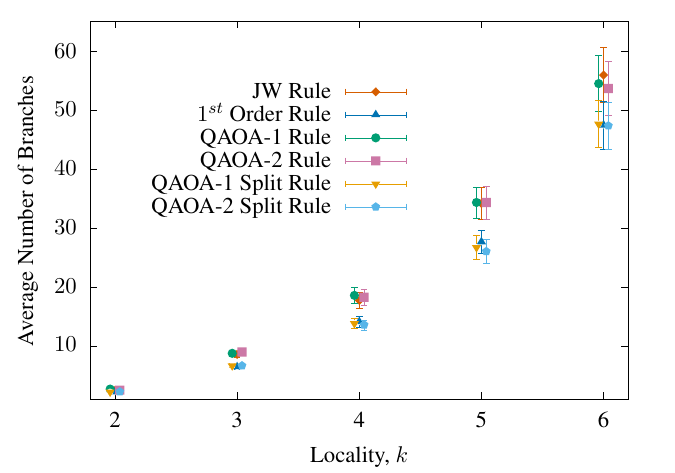}
\caption{The results of DPLL algorithms with various quantum and classical branching rules applied to random $k$-SAT problems with $n=8$ variables and $m = \lfloor n\ln(2) 2^k\rceil$ clauses.  The $y$-axis shows the average number of branching points that were needed by the DPLL algorithm to reach its conclusion.  Each point is averaged over 50 random instances, and the error bars represent the standard error of the mean.  This version includes the split version of QAOA meant to mimic the JW Rule.
}
\label{fig:DPLLsplit}
\end{figure}

For DPLL, we consider both classical branching rules as described by Eqs. (\ref{eq:JW}) \& (\ref{eq:firstorder}), the JW rule and the first-order rule respectively, and a quantum branching rule that uses QAOA-prepared expectation values.  Based on our analytics, we expect QAOA to roughly follow the performance the the first order classical branching rule, and numerically that is indeed what we see.

We show results from $n=15$ variable random $k$-SAT instances in Fig.~\ref{fig:DPLL15}.  The number of clauses is chosen to be $m = \lfloor n\ln(2) 2^k\rceil$ which is on the border between satisfiability and un-satisfiability, where the $k$-SAT problem is expected to be hard \cite{Achlioptas2003}.  The $y$-axis shows how many branches were needed for the algorithm to reach its conclusion.  We chose to count number of branches because this metric is agnostic of the details of the algorithm implementation which would currently vastly favor the classical branching rules due to the complexity of running simulations of quantum algorithms on classical computers.  Even comparing time from quantum hardware will depend heavily on the hardware itself and is not a good comparison of the base algorithms so much as the algorithms and hardware combined.

In practice we see that the JW rule does indeed outperform the first order or our quantum rules, with similar trends holding at other values of $n$.  Mostly, there is not enough statistical significance to distinguish the performance of QAOA versus the first order rules, but the first order rule does narrowly outperform QAOA.  Surprisingly QAOA-1 also narrowly outperforms QAOA-2 but without statistical significance.  This narrow detriment is likely due to the same effect that classically makes the JW rule outperform the first order rule \cite{Hooker1995}.  If we are selecting the direction to branch that is most advantageous, we necessarily will have a disadvantageous branch if we are forced to back-track.  Better performing rules, such as JW, lack symmetry as much, resulting in rules that are more balanced between the original branch and back-tracking.

This does raise the question of whether a quantum algorithm 
designed to emulate the JW rule, such as the one in Section~\ref{ssec:classical-inspired} would be able to track the performance of the JW rule.  These results are shown in Fig.~\ref{fig:DPLLsplit} where the classical-inspired quantum branching rules are referred to with the term ``split.''  These split rules do indeed track the JW rule, but note that these results are for $n=8$ qubits because the split version of QAOA takes twice as many qubits, reducing the system sizes we can simulate.  There is some minor performance improvement of these split rules over the classical JW rule, but these differences are not statistically significant and their extrapolation to larger system sizes is suspect.  Nevertheless, this indicates that we can indeed reverse engineer quantum branching rules that to perturbative order can mimic the behavior of classical rules.

\section{Conclusion}
\label{sec:conclusion}

This work has shown how quantum algorithms may be directly integrated into powerful classical frameworks, with the potential to yield greater computational advantage than by employing either component alone. Iterative quantum algorithms build off of a rich history of iterative classical algorithms and heuristics, and our general approach for dealing with constraints and high-order terms facilitates applications to a much wider variety of industrially-relevant optimization problems. Furthermore, we have shown how quantum resources may be integrated into classical algorithms beyond polynomial-time that perform complete search through our development of the Quantum DPLL solver.  The same ideas may be directly applied to other powerful classical search techniques, including branch-and-bound which we explore in forthcoming work.

Notably, we also provided tools for analyzing these quantum subcomponents in a general way to determine how they will behave in the larger algorithm in limiting cases.  The goal here is not just to slot in a quantum algorithm but to understand the quantum algorithm enough that we can predict and engineer its behavior, as well as help direct the design of future competitive heuristics.

The development of more advanced selection and reduction rules, leads to algorithms that are performing better than just using simple single bit selection and reduction rules.  Utilizing specific knowledge of the problem structure seems to lead to advantage, but the single bit fixing rules are usually more efficient to implement and lead to faster problem size reduction.  These more advanced techniques are also useful in dealing with hard constraints in problems, which are notoriously difficult to engineer into quantum algorithms.

Furthermore, our limiting behavior rules for small circuit depth QAOA and small $\gamma$ limits provides a general way of approaching the analytics of the quantum algorithms and providing a sense of how they will behave in the larger iterative framework.  In the case of DPLL, our limiting case analysis showed that the quantum algorithm would reproduce the behavior of a known classical algorithm, an insight that then allowed us to engineer a better quantum algorithm.  This general procedure can be used in most settings and problem classes.

With this toolset, we can find quantum algorithms that have low circuit depth limits that replicate the performance of existing classical algorithms.  That lets us engineer quantum algorithms at least to replicate the performance of classical algorithms.  If we can ensure analytically good performance from low depth quantum algorithms, the next step then needs to be finding ways to scale up the quantum algorithm that maintain and improve the performance. Ultimately this toolset seeks to advance more sophisticated hybrid paradigms yielding advantages over utilizing the quantum or classical components alone.

\acknowledgements

We thank our colleagues from NASA QuAIL for numerous helpful discussions, in particular Filip Maciejewski, Zoe Gonzalez Izquierdo, Shon Grabbe, and Eleanor Rieffel.
This material is based upon work supported by the U.S. Department of Energy, Office of Science, Office of Advanced Scientific Computing Research under Award Number 89243024SSC000129.
S.~H. was supported under the Prime Contract No.~80ARC020D0010 with the NASA Ames Research Center.

\bibliography{refs}

\begin{thebibliography}{44}%
\makeatletter
\providecommand \@ifxundefined [1]{%
 \@ifx{#1\undefined}
}%
\providecommand \@ifnum [1]{%
 \ifnum #1\expandafter \@firstoftwo
 \else \expandafter \@secondoftwo
 \fi
}%
\providecommand \@ifx [1]{%
 \ifx #1\expandafter \@firstoftwo
 \else \expandafter \@secondoftwo
 \fi
}%
\providecommand \natexlab [1]{#1}%
\providecommand \enquote  [1]{``#1''}%
\providecommand \bibnamefont  [1]{#1}%
\providecommand \bibfnamefont [1]{#1}%
\providecommand \citenamefont [1]{#1}%
\providecommand \href@noop [0]{\@secondoftwo}%
\providecommand \href [0]{\begingroup \@sanitize@url \@href}%
\providecommand \@href[1]{\@@startlink{#1}\@@href}%
\providecommand \@@href[1]{\endgroup#1\@@endlink}%
\providecommand \@sanitize@url [0]{\catcode `\\12\catcode `\$12\catcode `\&12\catcode `\#12\catcode `\^12\catcode `\_12\catcode `\%12\relax}%
\providecommand \@@startlink[1]{}%
\providecommand \@@endlink[0]{}%
\providecommand \url  [0]{\begingroup\@sanitize@url \@url }%
\providecommand \@url [1]{\endgroup\@href {#1}{\urlprefix }}%
\providecommand \urlprefix  [0]{URL }%
\providecommand \Eprint [0]{\href }%
\providecommand \doibase [0]{https://doi.org/}%
\providecommand \selectlanguage [0]{\@gobble}%
\providecommand \bibinfo  [0]{\@secondoftwo}%
\providecommand \bibfield  [0]{\@secondoftwo}%
\providecommand \translation [1]{[#1]}%
\providecommand \BibitemOpen [0]{}%
\providecommand \bibitemStop [0]{}%
\providecommand \bibitemNoStop [0]{.\EOS\space}%
\providecommand \EOS [0]{\spacefactor3000\relax}%
\providecommand \BibitemShut  [1]{\csname bibitem#1\endcsname}%
\let\auto@bib@innerbib\@empty
\bibitem [{\citenamefont {Bravyi}\ \emph {et~al.}(2020)\citenamefont {Bravyi}, \citenamefont {Kliesch}, \citenamefont {Koenig},\ and\ \citenamefont {Tang}}]{Bravyi2020}%
  \BibitemOpen
  \bibfield  {author} {\bibinfo {author} {\bibfnamefont {S.}~\bibnamefont {Bravyi}}, \bibinfo {author} {\bibfnamefont {A.}~\bibnamefont {Kliesch}}, \bibinfo {author} {\bibfnamefont {R.}~\bibnamefont {Koenig}},\ and\ \bibinfo {author} {\bibfnamefont {E.}~\bibnamefont {Tang}},\ }\bibfield  {title} {\bibinfo {title} {Obstacles to variational quantum optimization from symmetry protection},\ }\href {https://doi.org/10.1103/PhysRevLett.125.260505} {\bibfield  {journal} {\bibinfo  {journal} {Phys. Rev. Lett.}\ }\textbf {\bibinfo {volume} {125}},\ \bibinfo {pages} {260505} (\bibinfo {year} {2020})}\BibitemShut {NoStop}%
\bibitem [{\citenamefont {Dupont}\ \emph {et~al.}(2023)\citenamefont {Dupont}, \citenamefont {Evert}, \citenamefont {Hodson}, \citenamefont {Sundar}, \citenamefont {Jeffrey}, \citenamefont {Yamaguchi}, \citenamefont {Feng}, \citenamefont {Maciejewski}, \citenamefont {Hadfield}, \citenamefont {Alam}, \citenamefont {Wang}, \citenamefont {Grabbe}, \citenamefont {Lott}, \citenamefont {Rieffel}, \citenamefont {Venturelli},\ and\ \citenamefont {Reagor}}]{dupont2023quantum}%
  \BibitemOpen
  \bibfield  {author} {\bibinfo {author} {\bibfnamefont {M.}~\bibnamefont {Dupont}}, \bibinfo {author} {\bibfnamefont {B.}~\bibnamefont {Evert}}, \bibinfo {author} {\bibfnamefont {M.~J.}\ \bibnamefont {Hodson}}, \bibinfo {author} {\bibfnamefont {B.}~\bibnamefont {Sundar}}, \bibinfo {author} {\bibfnamefont {S.}~\bibnamefont {Jeffrey}}, \bibinfo {author} {\bibfnamefont {Y.}~\bibnamefont {Yamaguchi}}, \bibinfo {author} {\bibfnamefont {D.}~\bibnamefont {Feng}}, \bibinfo {author} {\bibfnamefont {F.~B.}\ \bibnamefont {Maciejewski}}, \bibinfo {author} {\bibfnamefont {S.}~\bibnamefont {Hadfield}}, \bibinfo {author} {\bibfnamefont {M.~S.}\ \bibnamefont {Alam}}, \bibinfo {author} {\bibfnamefont {Z.}~\bibnamefont {Wang}}, \bibinfo {author} {\bibfnamefont {S.}~\bibnamefont {Grabbe}}, \bibinfo {author} {\bibfnamefont {P.~A.}\ \bibnamefont {Lott}}, \bibinfo {author} {\bibfnamefont {E.~G.}\ \bibnamefont {Rieffel}}, \bibinfo {author} {\bibfnamefont {D.}~\bibnamefont {Venturelli}},\ and\ \bibinfo {author} {\bibfnamefont
  {M.~J.}\ \bibnamefont {Reagor}},\ }\bibfield  {title} {\bibinfo {title} {Quantum-enhanced greedy combinatorial optimization solver},\ }\href {https://doi.org/10.1126/sciadv.adi0487} {\bibfield  {journal} {\bibinfo  {journal} {Science Advances}\ }\textbf {\bibinfo {volume} {9}},\ \bibinfo {pages} {eadi0487} (\bibinfo {year} {2023})}\BibitemShut {NoStop}%
\bibitem [{\citenamefont {Brady}\ and\ \citenamefont {Hadfield}(2023)}]{Brady2023}%
  \BibitemOpen
  \bibfield  {author} {\bibinfo {author} {\bibfnamefont {L.~T.}\ \bibnamefont {Brady}}\ and\ \bibinfo {author} {\bibfnamefont {S.}~\bibnamefont {Hadfield}},\ }\href {https://arxiv.org/abs/2309.13110} {\bibinfo {title} {Iterative quantum algorithms for maximum independent set: A tale of low-depth quantum algorithms}} (\bibinfo {year} {2023}),\ \Eprint {https://arxiv.org/abs/2309.13110} {arXiv:2309.13110 [quant-ph]} \BibitemShut {NoStop}%
\bibitem [{\citenamefont {Finzgar}\ \emph {et~al.}(2023)\citenamefont {Finzgar}, \citenamefont {Kerschbaumer}, \citenamefont {Schuetz}, \citenamefont {Mendl},\ and\ \citenamefont {Katzgraber}}]{Finzgar2023}%
  \BibitemOpen
  \bibfield  {author} {\bibinfo {author} {\bibfnamefont {J.~R.}\ \bibnamefont {Finzgar}}, \bibinfo {author} {\bibfnamefont {A.}~\bibnamefont {Kerschbaumer}}, \bibinfo {author} {\bibfnamefont {M.~J.~A.}\ \bibnamefont {Schuetz}}, \bibinfo {author} {\bibfnamefont {C.~B.}\ \bibnamefont {Mendl}},\ and\ \bibinfo {author} {\bibfnamefont {H.~G.}\ \bibnamefont {Katzgraber}},\ }\href@noop {} {\bibinfo {title} {Quantum-informed recursive optimization algorithms}} (\bibinfo {year} {2023}),\ \Eprint {https://arxiv.org/abs/2308.13607} {arXiv:2308.13607 [quant-ph]} \BibitemShut {NoStop}%
\bibitem [{\citenamefont {Kadowaki}\ and\ \citenamefont {Nishimori}(1998)}]{Kadowaki_1998}%
  \BibitemOpen
  \bibfield  {author} {\bibinfo {author} {\bibfnamefont {T.}~\bibnamefont {Kadowaki}}\ and\ \bibinfo {author} {\bibfnamefont {H.}~\bibnamefont {Nishimori}},\ }\bibfield  {title} {\bibinfo {title} {Quantum annealing in the transverse ising model},\ }\href {https://doi.org/10.1103/physreve.58.5355} {\bibfield  {journal} {\bibinfo  {journal} {Physical Review E}\ }\textbf {\bibinfo {volume} {58}},\ \bibinfo {pages} {5355–5363} (\bibinfo {year} {1998})}\BibitemShut {NoStop}%
\bibitem [{\citenamefont {Farhi}\ \emph {et~al.}(2000)\citenamefont {Farhi}, \citenamefont {Goldstone}, \citenamefont {Gutmann},\ and\ \citenamefont {Sipser}}]{farhi2000}%
  \BibitemOpen
  \bibfield  {author} {\bibinfo {author} {\bibfnamefont {E.}~\bibnamefont {Farhi}}, \bibinfo {author} {\bibfnamefont {J.}~\bibnamefont {Goldstone}}, \bibinfo {author} {\bibfnamefont {S.}~\bibnamefont {Gutmann}},\ and\ \bibinfo {author} {\bibfnamefont {M.}~\bibnamefont {Sipser}},\ }\href {https://arxiv.org/abs/quant-ph/0001106} {\bibinfo {title} {Quantum computation by adiabatic evolution}} (\bibinfo {year} {2000}),\ \Eprint {https://arxiv.org/abs/quant-ph/0001106} {arXiv:quant-ph/0001106 [quant-ph]} \BibitemShut {NoStop}%
\bibitem [{\citenamefont {Crosson}\ and\ \citenamefont {Lidar}(2021)}]{Crosson_2021}%
  \BibitemOpen
  \bibfield  {author} {\bibinfo {author} {\bibfnamefont {E.~J.}\ \bibnamefont {Crosson}}\ and\ \bibinfo {author} {\bibfnamefont {D.~A.}\ \bibnamefont {Lidar}},\ }\bibfield  {title} {\bibinfo {title} {Prospects for quantum enhancement with diabatic quantum annealing},\ }\href {https://doi.org/10.1038/s42254-021-00313-6} {\bibfield  {journal} {\bibinfo  {journal} {Nature Reviews Physics}\ }\textbf {\bibinfo {volume} {3}},\ \bibinfo {pages} {466–489} (\bibinfo {year} {2021})}\BibitemShut {NoStop}%
\bibitem [{\citenamefont {Farhi}\ \emph {et~al.}(2014)\citenamefont {Farhi}, \citenamefont {Goldstone},\ and\ \citenamefont {Gutmann}}]{Farhi2014}%
  \BibitemOpen
  \bibfield  {author} {\bibinfo {author} {\bibfnamefont {E.}~\bibnamefont {Farhi}}, \bibinfo {author} {\bibfnamefont {J.}~\bibnamefont {Goldstone}},\ and\ \bibinfo {author} {\bibfnamefont {S.}~\bibnamefont {Gutmann}},\ }\href {https://doi.org/10.48550/ARXIV.1411.4028} {\bibinfo {title} {A quantum approximate optimization algorithm}} (\bibinfo {year} {2014})\BibitemShut {NoStop}%
\bibitem [{\citenamefont {Hadfield}\ \emph {et~al.}(2019)\citenamefont {Hadfield}, \citenamefont {Wang}, \citenamefont {O’Gorman}, \citenamefont {Rieffel}, \citenamefont {Venturelli},\ and\ \citenamefont {Biswas}}]{hadfield2019quantum}%
  \BibitemOpen
  \bibfield  {author} {\bibinfo {author} {\bibfnamefont {S.}~\bibnamefont {Hadfield}}, \bibinfo {author} {\bibfnamefont {Z.}~\bibnamefont {Wang}}, \bibinfo {author} {\bibfnamefont {B.}~\bibnamefont {O’Gorman}}, \bibinfo {author} {\bibfnamefont {E.~G.}\ \bibnamefont {Rieffel}}, \bibinfo {author} {\bibfnamefont {D.}~\bibnamefont {Venturelli}},\ and\ \bibinfo {author} {\bibfnamefont {R.}~\bibnamefont {Biswas}},\ }\bibfield  {title} {\bibinfo {title} {From the quantum approximate optimization algorithm to a quantum alternating operator ansatz},\ }\href@noop {} {\bibfield  {journal} {\bibinfo  {journal} {Algorithms}\ }\textbf {\bibinfo {volume} {12}},\ \bibinfo {pages} {34} (\bibinfo {year} {2019})}\BibitemShut {NoStop}%
\bibitem [{\citenamefont {Peruzzo}\ \emph {et~al.}(2014)\citenamefont {Peruzzo}, \citenamefont {McClean}, \citenamefont {Shadbolt}, \citenamefont {Yung}, \citenamefont {Zhou}, \citenamefont {Love}, \citenamefont {Aspuru-Guzik},\ and\ \citenamefont {O'Brien}}]{Peruzzo2014}%
  \BibitemOpen
  \bibfield  {author} {\bibinfo {author} {\bibfnamefont {A.}~\bibnamefont {Peruzzo}}, \bibinfo {author} {\bibfnamefont {J.}~\bibnamefont {McClean}}, \bibinfo {author} {\bibfnamefont {P.}~\bibnamefont {Shadbolt}}, \bibinfo {author} {\bibfnamefont {M.-H.}\ \bibnamefont {Yung}}, \bibinfo {author} {\bibfnamefont {X.-Q.}\ \bibnamefont {Zhou}}, \bibinfo {author} {\bibfnamefont {P.~J.}\ \bibnamefont {Love}}, \bibinfo {author} {\bibfnamefont {A.}~\bibnamefont {Aspuru-Guzik}},\ and\ \bibinfo {author} {\bibfnamefont {J.~L.}\ \bibnamefont {O'Brien}},\ }\bibfield  {title} {\bibinfo {title} {A variational eigenvalue solver on a photonic quantum processor},\ }\bibfield  {journal} {\bibinfo  {journal} {Nature Communications}\ }\textbf {\bibinfo {volume} {5}},\ \href {https://doi.org/10.1038/ncomms5213} {10.1038/ncomms5213} (\bibinfo {year} {2014})\BibitemShut {NoStop}%
\bibitem [{\citenamefont {Cerezo}\ \emph {et~al.}(2021)\citenamefont {Cerezo}, \citenamefont {Arrasmith}, \citenamefont {Babbush}, \citenamefont {Benjamin}, \citenamefont {Endo}, \citenamefont {Fujii}, \citenamefont {McClean}, \citenamefont {Mitarai}, \citenamefont {Yuan}, \citenamefont {Cincio} \emph {et~al.}}]{Cerezo2021}%
  \BibitemOpen
  \bibfield  {author} {\bibinfo {author} {\bibfnamefont {M.}~\bibnamefont {Cerezo}}, \bibinfo {author} {\bibfnamefont {A.}~\bibnamefont {Arrasmith}}, \bibinfo {author} {\bibfnamefont {R.}~\bibnamefont {Babbush}}, \bibinfo {author} {\bibfnamefont {S.~C.}\ \bibnamefont {Benjamin}}, \bibinfo {author} {\bibfnamefont {S.}~\bibnamefont {Endo}}, \bibinfo {author} {\bibfnamefont {K.}~\bibnamefont {Fujii}}, \bibinfo {author} {\bibfnamefont {J.~R.}\ \bibnamefont {McClean}}, \bibinfo {author} {\bibfnamefont {K.}~\bibnamefont {Mitarai}}, \bibinfo {author} {\bibfnamefont {X.}~\bibnamefont {Yuan}}, \bibinfo {author} {\bibfnamefont {L.}~\bibnamefont {Cincio}}, \emph {et~al.},\ }\bibfield  {title} {\bibinfo {title} {Variational quantum algorithms},\ }\href {https://doi.org/https://doi.org/10.1038/s42254-021-00348-9} {\bibfield  {journal} {\bibinfo  {journal} {Nature Reviews Physics}\ }\textbf {\bibinfo {volume} {3}},\ \bibinfo {pages} {625} (\bibinfo {year} {2021})}\BibitemShut {NoStop}%
\bibitem [{\citenamefont {Maciejewski}\ \emph {et~al.}(2024)\citenamefont {Maciejewski}, \citenamefont {Hadfield}, \citenamefont {Hall}, \citenamefont {Hodson}, \citenamefont {Dupont}, \citenamefont {Evert}, \citenamefont {Sud}, \citenamefont {Alam}, \citenamefont {Wang}, \citenamefont {Jeffrey}, \citenamefont {Sundar}, \citenamefont {Lott}, \citenamefont {Grabbe}, \citenamefont {Rieffel}, \citenamefont {Reagor},\ and\ \citenamefont {Venturelli}}]{Maciejewski2024}%
  \BibitemOpen
  \bibfield  {author} {\bibinfo {author} {\bibfnamefont {F.~B.}\ \bibnamefont {Maciejewski}}, \bibinfo {author} {\bibfnamefont {S.}~\bibnamefont {Hadfield}}, \bibinfo {author} {\bibfnamefont {B.}~\bibnamefont {Hall}}, \bibinfo {author} {\bibfnamefont {M.}~\bibnamefont {Hodson}}, \bibinfo {author} {\bibfnamefont {M.}~\bibnamefont {Dupont}}, \bibinfo {author} {\bibfnamefont {B.}~\bibnamefont {Evert}}, \bibinfo {author} {\bibfnamefont {J.}~\bibnamefont {Sud}}, \bibinfo {author} {\bibfnamefont {M.~S.}\ \bibnamefont {Alam}}, \bibinfo {author} {\bibfnamefont {Z.}~\bibnamefont {Wang}}, \bibinfo {author} {\bibfnamefont {S.}~\bibnamefont {Jeffrey}}, \bibinfo {author} {\bibfnamefont {B.}~\bibnamefont {Sundar}}, \bibinfo {author} {\bibfnamefont {P.~A.}\ \bibnamefont {Lott}}, \bibinfo {author} {\bibfnamefont {S.}~\bibnamefont {Grabbe}}, \bibinfo {author} {\bibfnamefont {E.~G.}\ \bibnamefont {Rieffel}}, \bibinfo {author} {\bibfnamefont {M.~J.}\ \bibnamefont {Reagor}},\ and\ \bibinfo {author} {\bibfnamefont
  {D.}~\bibnamefont {Venturelli}},\ }\href {https://arxiv.org/abs/2308.12423} {\bibinfo {title} {Design and execution of quantum circuits using tens of superconducting qubits and thousands of gates for dense ising optimization problems}} (\bibinfo {year} {2024}),\ \Eprint {https://arxiv.org/abs/2308.12423} {arXiv:2308.12423 [quant-ph]} \BibitemShut {NoStop}%
\bibitem [{\citenamefont {Bravyi}\ \emph {et~al.}(2022)\citenamefont {Bravyi}, \citenamefont {Kliesch}, \citenamefont {Koenig},\ and\ \citenamefont {Tang}}]{Bravyi2022hybridquantum}%
  \BibitemOpen
  \bibfield  {author} {\bibinfo {author} {\bibfnamefont {S.}~\bibnamefont {Bravyi}}, \bibinfo {author} {\bibfnamefont {A.}~\bibnamefont {Kliesch}}, \bibinfo {author} {\bibfnamefont {R.}~\bibnamefont {Koenig}},\ and\ \bibinfo {author} {\bibfnamefont {E.}~\bibnamefont {Tang}},\ }\bibfield  {title} {\bibinfo {title} {Hybrid quantum-classical algorithms for approximate graph coloring},\ }\href {https://doi.org/10.22331/q-2022-03-30-678} {\bibfield  {journal} {\bibinfo  {journal} {{Quantum}}\ }\textbf {\bibinfo {volume} {6}},\ \bibinfo {pages} {678} (\bibinfo {year} {2022})}\BibitemShut {NoStop}%
\bibitem [{\citenamefont {Bittel}\ and\ \citenamefont {Kliesch}(2021)}]{Bittel2021}%
  \BibitemOpen
  \bibfield  {author} {\bibinfo {author} {\bibfnamefont {L.}~\bibnamefont {Bittel}}\ and\ \bibinfo {author} {\bibfnamefont {M.}~\bibnamefont {Kliesch}},\ }\bibfield  {title} {\bibinfo {title} {Training variational quantum algorithms is {NP}-hard},\ }\bibfield  {journal} {\bibinfo  {journal} {Physical Review Letters}\ }\textbf {\bibinfo {volume} {127}},\ \href {https://doi.org/10.1103/physrevlett.127.120502} {10.1103/physrevlett.127.120502} (\bibinfo {year} {2021})\BibitemShut {NoStop}%
\bibitem [{\citenamefont {Ragone}\ \emph {et~al.}(2024)\citenamefont {Ragone}, \citenamefont {Bakalov}, \citenamefont {Sauvage}, \citenamefont {Kemper}, \citenamefont {Ortiz~Marrero}, \citenamefont {Larocca},\ and\ \citenamefont {Cerezo}}]{Ragone_2024}%
  \BibitemOpen
  \bibfield  {author} {\bibinfo {author} {\bibfnamefont {M.}~\bibnamefont {Ragone}}, \bibinfo {author} {\bibfnamefont {B.~N.}\ \bibnamefont {Bakalov}}, \bibinfo {author} {\bibfnamefont {F.}~\bibnamefont {Sauvage}}, \bibinfo {author} {\bibfnamefont {A.~F.}\ \bibnamefont {Kemper}}, \bibinfo {author} {\bibfnamefont {C.}~\bibnamefont {Ortiz~Marrero}}, \bibinfo {author} {\bibfnamefont {M.}~\bibnamefont {Larocca}},\ and\ \bibinfo {author} {\bibfnamefont {M.}~\bibnamefont {Cerezo}},\ }\bibfield  {title} {\bibinfo {title} {A lie algebraic theory of barren plateaus for deep parameterized quantum circuits},\ }\bibfield  {journal} {\bibinfo  {journal} {Nature Communications}\ }\textbf {\bibinfo {volume} {15}},\ \href {https://doi.org/10.1038/s41467-024-49909-3} {10.1038/s41467-024-49909-3} (\bibinfo {year} {2024})\BibitemShut {NoStop}%
\bibitem [{\citenamefont {Fontana}\ \emph {et~al.}(2024)\citenamefont {Fontana}, \citenamefont {Herman}, \citenamefont {Chakrabarti}, \citenamefont {Kumar}, \citenamefont {Yalovetzky}, \citenamefont {Heredge}, \citenamefont {Sureshbabu},\ and\ \citenamefont {Pistoia}}]{Fontana_2024}%
  \BibitemOpen
  \bibfield  {author} {\bibinfo {author} {\bibfnamefont {E.}~\bibnamefont {Fontana}}, \bibinfo {author} {\bibfnamefont {D.}~\bibnamefont {Herman}}, \bibinfo {author} {\bibfnamefont {S.}~\bibnamefont {Chakrabarti}}, \bibinfo {author} {\bibfnamefont {N.}~\bibnamefont {Kumar}}, \bibinfo {author} {\bibfnamefont {R.}~\bibnamefont {Yalovetzky}}, \bibinfo {author} {\bibfnamefont {J.}~\bibnamefont {Heredge}}, \bibinfo {author} {\bibfnamefont {S.~H.}\ \bibnamefont {Sureshbabu}},\ and\ \bibinfo {author} {\bibfnamefont {M.}~\bibnamefont {Pistoia}},\ }\bibfield  {title} {\bibinfo {title} {Characterizing barren plateaus in quantum ansätze with the adjoint representation},\ }\bibfield  {journal} {\bibinfo  {journal} {Nature Communications}\ }\textbf {\bibinfo {volume} {15}},\ \href {https://doi.org/10.1038/s41467-024-49910-w} {10.1038/s41467-024-49910-w} (\bibinfo {year} {2024})\BibitemShut {NoStop}%
\bibitem [{\citenamefont {Akshay}\ \emph {et~al.}(2021)\citenamefont {Akshay}, \citenamefont {Rabinovich}, \citenamefont {Campos},\ and\ \citenamefont {Biamonte}}]{Akshay_2021}%
  \BibitemOpen
  \bibfield  {author} {\bibinfo {author} {\bibfnamefont {V.}~\bibnamefont {Akshay}}, \bibinfo {author} {\bibfnamefont {D.}~\bibnamefont {Rabinovich}}, \bibinfo {author} {\bibfnamefont {E.}~\bibnamefont {Campos}},\ and\ \bibinfo {author} {\bibfnamefont {J.}~\bibnamefont {Biamonte}},\ }\bibfield  {title} {\bibinfo {title} {Parameter concentrations in quantum approximate optimization},\ }\bibfield  {journal} {\bibinfo  {journal} {Physical Review A}\ }\textbf {\bibinfo {volume} {104}},\ \href {https://doi.org/10.1103/physreva.104.l010401} {10.1103/physreva.104.l010401} (\bibinfo {year} {2021})\BibitemShut {NoStop}%
\bibitem [{\citenamefont {Wurtz}\ and\ \citenamefont {Lykov}(2021)}]{Wurtz_2021}%
  \BibitemOpen
  \bibfield  {author} {\bibinfo {author} {\bibfnamefont {J.}~\bibnamefont {Wurtz}}\ and\ \bibinfo {author} {\bibfnamefont {D.}~\bibnamefont {Lykov}},\ }\bibfield  {title} {\bibinfo {title} {Fixed-angle conjectures for the quantum approximate optimization algorithm on regular maxcut graphs},\ }\href {https://doi.org/10.1103/PhysRevA.104.052419} {\bibfield  {journal} {\bibinfo  {journal} {Phys. Rev. A}\ }\textbf {\bibinfo {volume} {104}},\ \bibinfo {pages} {052419} (\bibinfo {year} {2021})}\BibitemShut {NoStop}%
\bibitem [{\citenamefont {Shaydulin}\ \emph {et~al.}(2023)\citenamefont {Shaydulin}, \citenamefont {Lotshaw}, \citenamefont {Larson}, \citenamefont {Ostrowski},\ and\ \citenamefont {Humble}}]{Shaydulin_2023}%
  \BibitemOpen
  \bibfield  {author} {\bibinfo {author} {\bibfnamefont {R.}~\bibnamefont {Shaydulin}}, \bibinfo {author} {\bibfnamefont {P.~C.}\ \bibnamefont {Lotshaw}}, \bibinfo {author} {\bibfnamefont {J.}~\bibnamefont {Larson}}, \bibinfo {author} {\bibfnamefont {J.}~\bibnamefont {Ostrowski}},\ and\ \bibinfo {author} {\bibfnamefont {T.~S.}\ \bibnamefont {Humble}},\ }\bibfield  {title} {\bibinfo {title} {Parameter transfer for quantum approximate optimization of weighted maxcut},\ }\href {https://doi.org/10.1145/3584706} {\bibfield  {journal} {\bibinfo  {journal} {ACM Transactions on Quantum Computing}\ }\textbf {\bibinfo {volume} {4}},\ \bibinfo {pages} {1–15} (\bibinfo {year} {2023})}\BibitemShut {NoStop}%
\bibitem [{\citenamefont {Magann}\ \emph {et~al.}(2022)\citenamefont {Magann}, \citenamefont {Rudinger}, \citenamefont {Grace},\ and\ \citenamefont {Sarovar}}]{Magann_2022}%
  \BibitemOpen
  \bibfield  {author} {\bibinfo {author} {\bibfnamefont {A.~B.}\ \bibnamefont {Magann}}, \bibinfo {author} {\bibfnamefont {K.~M.}\ \bibnamefont {Rudinger}}, \bibinfo {author} {\bibfnamefont {M.~D.}\ \bibnamefont {Grace}},\ and\ \bibinfo {author} {\bibfnamefont {M.}~\bibnamefont {Sarovar}},\ }\bibfield  {title} {\bibinfo {title} {Feedback-based quantum optimization},\ }\bibfield  {journal} {\bibinfo  {journal} {Physical Review Letters}\ }\textbf {\bibinfo {volume} {129}},\ \href {https://doi.org/10.1103/physrevlett.129.250502} {10.1103/physrevlett.129.250502} (\bibinfo {year} {2022})\BibitemShut {NoStop}%
\bibitem [{\citenamefont {Brady}\ and\ \citenamefont {Hadfield}(2024)}]{Brady_2024}%
  \BibitemOpen
  \bibfield  {author} {\bibinfo {author} {\bibfnamefont {L.~T.}\ \bibnamefont {Brady}}\ and\ \bibinfo {author} {\bibfnamefont {S.}~\bibnamefont {Hadfield}},\ }\href {https://arxiv.org/abs/2409.15426} {\bibinfo {title} {Focqs: Feedback optimally controlled quantum states}} (\bibinfo {year} {2024}),\ \Eprint {https://arxiv.org/abs/2409.15426} {arXiv:2409.15426 [quant-ph]} \BibitemShut {NoStop}%
\bibitem [{\citenamefont {Davis}\ and\ \citenamefont {Putnam}(1960)}]{Davis_1960}%
  \BibitemOpen
  \bibfield  {author} {\bibinfo {author} {\bibfnamefont {M.}~\bibnamefont {Davis}}\ and\ \bibinfo {author} {\bibfnamefont {H.}~\bibnamefont {Putnam}},\ }\bibfield  {title} {\bibinfo {title} {A computing procedure for quantification theory},\ }\href {https://doi.org/10.1145/321033.321034} {\bibfield  {journal} {\bibinfo  {journal} {J. ACM}\ }\textbf {\bibinfo {volume} {7}},\ \bibinfo {pages} {201–215} (\bibinfo {year} {1960})}\BibitemShut {NoStop}%
\bibitem [{\citenamefont {Davis}\ \emph {et~al.}(1962)\citenamefont {Davis}, \citenamefont {Logemann},\ and\ \citenamefont {Loveland}}]{Davis_1962}%
  \BibitemOpen
  \bibfield  {author} {\bibinfo {author} {\bibfnamefont {M.}~\bibnamefont {Davis}}, \bibinfo {author} {\bibfnamefont {G.}~\bibnamefont {Logemann}},\ and\ \bibinfo {author} {\bibfnamefont {D.}~\bibnamefont {Loveland}},\ }\bibfield  {title} {\bibinfo {title} {A machine program for theorem-proving},\ }\href {https://doi.org/10.1145/368273.368557} {\bibfield  {journal} {\bibinfo  {journal} {Commun. ACM}\ }\textbf {\bibinfo {volume} {5}},\ \bibinfo {pages} {394–397} (\bibinfo {year} {1962})}\BibitemShut {NoStop}%
\bibitem [{\citenamefont {Lucas}(2014)}]{lucas2014ising}%
  \BibitemOpen
  \bibfield  {author} {\bibinfo {author} {\bibfnamefont {A.}~\bibnamefont {Lucas}},\ }\bibfield  {title} {\bibinfo {title} {Ising formulations of many np problems},\ }\href@noop {} {\bibfield  {journal} {\bibinfo  {journal} {Frontiers in physics}\ }\textbf {\bibinfo {volume} {2}},\ \bibinfo {pages} {5} (\bibinfo {year} {2014})}\BibitemShut {NoStop}%
\bibitem [{\citenamefont {Hadfield}(2021)}]{hadfield2021representation}%
  \BibitemOpen
  \bibfield  {author} {\bibinfo {author} {\bibfnamefont {S.}~\bibnamefont {Hadfield}},\ }\bibfield  {title} {\bibinfo {title} {On the representation of boolean and real functions as hamiltonians for quantum computing},\ }\href@noop {} {\bibfield  {journal} {\bibinfo  {journal} {ACM Transactions on Quantum Computing}\ }\textbf {\bibinfo {volume} {2}},\ \bibinfo {pages} {1} (\bibinfo {year} {2021})}\BibitemShut {NoStop}%
\bibitem [{\citenamefont {Quine}(1952)}]{Quine01101952}%
  \BibitemOpen
  \bibfield  {author} {\bibinfo {author} {\bibfnamefont {W.~V.}\ \bibnamefont {Quine}},\ }\bibfield  {title} {\bibinfo {title} {The problem of simplifying truth functions},\ }\href {https://doi.org/10.1080/00029890.1952.11988183} {\bibfield  {journal} {\bibinfo  {journal} {The American Mathematical Monthly}\ }\textbf {\bibinfo {volume} {59}},\ \bibinfo {pages} {521} (\bibinfo {year} {1952})},\ \Eprint {https://arxiv.org/abs/https://doi.org/10.1080/00029890.1952.11988183} {https://doi.org/10.1080/00029890.1952.11988183} \BibitemShut {NoStop}%
\bibitem [{\citenamefont {Quine}(1955)}]{Quine01111955}%
  \BibitemOpen
  \bibfield  {author} {\bibinfo {author} {\bibfnamefont {W.~V.}\ \bibnamefont {Quine}},\ }\bibfield  {title} {\bibinfo {title} {A way to simplify truth functions},\ }\href {https://doi.org/10.1080/00029890.1955.11988710} {\bibfield  {journal} {\bibinfo  {journal} {The American Mathematical Monthly}\ }\textbf {\bibinfo {volume} {62}},\ \bibinfo {pages} {627} (\bibinfo {year} {1955})},\ \Eprint {https://arxiv.org/abs/https://doi.org/10.1080/00029890.1955.11988710} {https://doi.org/10.1080/00029890.1955.11988710} \BibitemShut {NoStop}%
\bibitem [{\citenamefont {McCluskey}(1956)}]{McCluskey1956}%
  \BibitemOpen
  \bibfield  {author} {\bibinfo {author} {\bibfnamefont {E.~J.}\ \bibnamefont {McCluskey}},\ }\bibfield  {title} {\bibinfo {title} {Minimization of boolean functions},\ }\href {https://doi.org/10.1002/j.1538-7305.1956.tb03835.x} {\bibfield  {journal} {\bibinfo  {journal} {The Bell System Technical Journal}\ }\textbf {\bibinfo {volume} {35}},\ \bibinfo {pages} {1417} (\bibinfo {year} {1956})}\BibitemShut {NoStop}%
\bibitem [{\citenamefont {Marques~Silva}\ and\ \citenamefont {Sakallah}(1996)}]{569607}%
  \BibitemOpen
  \bibfield  {author} {\bibinfo {author} {\bibfnamefont {J.}~\bibnamefont {Marques~Silva}}\ and\ \bibinfo {author} {\bibfnamefont {K.}~\bibnamefont {Sakallah}},\ }\bibfield  {title} {\bibinfo {title} {Grasp-a new search algorithm for satisfiability},\ }in\ \href {https://doi.org/10.1109/ICCAD.1996.569607} {\emph {\bibinfo {booktitle} {Proceedings of International Conference on Computer Aided Design}}}\ (\bibinfo {year} {1996})\ pp.\ \bibinfo {pages} {220--227}\BibitemShut {NoStop}%
\bibitem [{\citenamefont {Marques-Silva}\ and\ \citenamefont {Sakallah}(1999)}]{769433}%
  \BibitemOpen
  \bibfield  {author} {\bibinfo {author} {\bibfnamefont {J.}~\bibnamefont {Marques-Silva}}\ and\ \bibinfo {author} {\bibfnamefont {K.}~\bibnamefont {Sakallah}},\ }\bibfield  {title} {\bibinfo {title} {Grasp: a search algorithm for propositional satisfiability},\ }\href {https://doi.org/10.1109/12.769433} {\bibfield  {journal} {\bibinfo  {journal} {IEEE Transactions on Computers}\ }\textbf {\bibinfo {volume} {48}},\ \bibinfo {pages} {506} (\bibinfo {year} {1999})}\BibitemShut {NoStop}%
\bibitem [{\citenamefont {Bayardo}\ and\ \citenamefont {Schrag}(1997)}]{10.5555/1867406.1867438}%
  \BibitemOpen
  \bibfield  {author} {\bibinfo {author} {\bibfnamefont {R.~J.}\ \bibnamefont {Bayardo}}\ and\ \bibinfo {author} {\bibfnamefont {R.~C.}\ \bibnamefont {Schrag}},\ }\bibfield  {title} {\bibinfo {title} {Using csp look-back techniques to solve real-world sat instances},\ }in\ \href@noop {} {\emph {\bibinfo {booktitle} {Proceedings of the Fourteenth National Conference on Artificial Intelligence and Ninth Conference on Innovative Applications of Artificial Intelligence}}},\ \bibinfo {series and number} {AAAI'97/IAAI'97}\ (\bibinfo  {publisher} {AAAI Press},\ \bibinfo {year} {1997})\ p.\ \bibinfo {pages} {203–208}\BibitemShut {NoStop}%
\bibitem [{SAT()}]{SAT_Competition}%
  \BibitemOpen
  \href {https://satcompetition.github.io/} {\bibinfo {title} {The international sat competition web page}}\BibitemShut {NoStop}%
\bibitem [{\citenamefont {Biere}\ \emph {et~al.}(2024)\citenamefont {Biere}, \citenamefont {Faller}, \citenamefont {Fazekas}, \citenamefont {Fleury}, \citenamefont {Froleyks},\ and\ \citenamefont {Pollitt}}]{BiereFallerFazekasFleuryFroleyks-CAV24}%
  \BibitemOpen
  \bibfield  {author} {\bibinfo {author} {\bibfnamefont {A.}~\bibnamefont {Biere}}, \bibinfo {author} {\bibfnamefont {T.}~\bibnamefont {Faller}}, \bibinfo {author} {\bibfnamefont {K.}~\bibnamefont {Fazekas}}, \bibinfo {author} {\bibfnamefont {M.}~\bibnamefont {Fleury}}, \bibinfo {author} {\bibfnamefont {N.}~\bibnamefont {Froleyks}},\ and\ \bibinfo {author} {\bibfnamefont {F.}~\bibnamefont {Pollitt}},\ }\bibfield  {title} {\bibinfo {title} {{CaDiCaL 2.0}},\ }in\ \href {https://doi.org/10.1007/978-3-031-65627-9\_7} {\emph {\bibinfo {booktitle} {Computer Aided Verification - 36th International Conference, {CAV} 2024, Montreal, QC, Canada, July 24-27, 2024, Proceedings, Part {I}}}},\ \bibinfo {series} {Lecture Notes in Computer Science}, Vol.\ \bibinfo {volume} {14681},\ \bibinfo {editor} {edited by\ \bibinfo {editor} {\bibfnamefont {A.}~\bibnamefont {Gurfinkel}}\ and\ \bibinfo {editor} {\bibfnamefont {V.}~\bibnamefont {Ganesh}}}\ (\bibinfo  {publisher} {Springer},\ \bibinfo {year} {2024})\ pp.\ \bibinfo {pages}
  {133--152}\BibitemShut {NoStop}%
\bibitem [{\citenamefont {Hooker}\ \emph {et~al.}(1991)\citenamefont {Hooker}, \citenamefont {harche},\ and\ \citenamefont {Thompson}}]{Hooker_1991}%
  \BibitemOpen
  \bibfield  {author} {\bibinfo {author} {\bibfnamefont {J.}~\bibnamefont {Hooker}}, \bibinfo {author} {\bibfnamefont {F.}~\bibnamefont {harche}},\ and\ \bibinfo {author} {\bibfnamefont {G.}~\bibnamefont {Thompson}},\ }\href {https://ideas.repec.org/p/cmu/gsiawp/1991-27.html} {\emph {\bibinfo {title} {{A Computational Stufy of Satisfiability Algorithms for Propositional Logic}}}},\ \bibinfo {type} {GSIA Working Papers}\ \bibinfo {number} {1991-27}\ (\bibinfo  {institution} {Carnegie Mellon University, Tepper School of Business},\ \bibinfo {year} {1991})\BibitemShut {NoStop}%
\bibitem [{\citenamefont {Hooker}\ and\ \citenamefont {Vinay}(1995)}]{Hooker1995}%
  \BibitemOpen
  \bibfield  {author} {\bibinfo {author} {\bibfnamefont {J.~N.}\ \bibnamefont {Hooker}}\ and\ \bibinfo {author} {\bibfnamefont {V.}~\bibnamefont {Vinay}},\ }\bibfield  {title} {\bibinfo {title} {Branching rules for satisfiability},\ }\href {https://doi.org/10.1007/BF00881805} {\bibfield  {journal} {\bibinfo  {journal} {J Autom Reasoning}\ ,\ \bibinfo {pages} {359–383}} (\bibinfo {year} {1995})}\BibitemShut {NoStop}%
\bibitem [{\citenamefont {Montanaro}(2016)}]{Montanaro2016}%
  \BibitemOpen
  \bibfield  {author} {\bibinfo {author} {\bibfnamefont {A.}~\bibnamefont {Montanaro}},\ }\href {https://arxiv.org/abs/1509.02374} {\bibinfo {title} {Quantum walk speedup of backtracking algorithms}} (\bibinfo {year} {2016}),\ \Eprint {https://arxiv.org/abs/1509.02374} {arXiv:1509.02374 [quant-ph]} \BibitemShut {NoStop}%
\bibitem [{\citenamefont {Hadfield}\ \emph {et~al.}(2022)\citenamefont {Hadfield}, \citenamefont {Hogg},\ and\ \citenamefont {Rieffel}}]{hadfield2022analytical}%
  \BibitemOpen
  \bibfield  {author} {\bibinfo {author} {\bibfnamefont {S.}~\bibnamefont {Hadfield}}, \bibinfo {author} {\bibfnamefont {T.}~\bibnamefont {Hogg}},\ and\ \bibinfo {author} {\bibfnamefont {E.~G.}\ \bibnamefont {Rieffel}},\ }\bibfield  {title} {\bibinfo {title} {Analytical framework for quantum alternating operator ans{\"a}tze},\ }\href@noop {} {\bibfield  {journal} {\bibinfo  {journal} {Quantum Science and Technology}\ }\textbf {\bibinfo {volume} {8}},\ \bibinfo {pages} {015017} (\bibinfo {year} {2022})}\BibitemShut {NoStop}%
\bibitem [{\citenamefont {Wurtz}\ and\ \citenamefont {Love}(2022)}]{Wurtz_2022}%
  \BibitemOpen
  \bibfield  {author} {\bibinfo {author} {\bibfnamefont {J.}~\bibnamefont {Wurtz}}\ and\ \bibinfo {author} {\bibfnamefont {P.~J.}\ \bibnamefont {Love}},\ }\bibfield  {title} {\bibinfo {title} {Counterdiabaticity and the quantum approximate optimization algorithm},\ }\href {https://doi.org/10.22331/q-2022-01-27-635} {\bibfield  {journal} {\bibinfo  {journal} {Quantum}\ }\textbf {\bibinfo {volume} {6}},\ \bibinfo {pages} {635} (\bibinfo {year} {2022})}\BibitemShut {NoStop}%
\bibitem [{\citenamefont {Zhou}\ \emph {et~al.}(2020)\citenamefont {Zhou}, \citenamefont {Wang}, \citenamefont {Choi}, \citenamefont {Pichler},\ and\ \citenamefont {Lukin}}]{Zhou_2020}%
  \BibitemOpen
  \bibfield  {author} {\bibinfo {author} {\bibfnamefont {L.}~\bibnamefont {Zhou}}, \bibinfo {author} {\bibfnamefont {S.-T.}\ \bibnamefont {Wang}}, \bibinfo {author} {\bibfnamefont {S.}~\bibnamefont {Choi}}, \bibinfo {author} {\bibfnamefont {H.}~\bibnamefont {Pichler}},\ and\ \bibinfo {author} {\bibfnamefont {M.~D.}\ \bibnamefont {Lukin}},\ }\bibfield  {title} {\bibinfo {title} {Quantum approximate optimization algorithm: Performance, mechanism, and implementation on near-term devices},\ }\bibfield  {journal} {\bibinfo  {journal} {Physical Review X}\ }\textbf {\bibinfo {volume} {10}},\ \href {https://doi.org/10.1103/physrevx.10.021067} {10.1103/physrevx.10.021067} (\bibinfo {year} {2020})\BibitemShut {NoStop}%
\bibitem [{\citenamefont {Pagano}\ \emph {et~al.}(2020)\citenamefont {Pagano}, \citenamefont {Bapat}, \citenamefont {Becker}, \citenamefont {Collins}, \citenamefont {De}, \citenamefont {Hess}, \citenamefont {Kaplan}, \citenamefont {Kyprianidis}, \citenamefont {Tan}, \citenamefont {Baldwin}, \citenamefont {Brady}, \citenamefont {Deshpande}, \citenamefont {Liu}, \citenamefont {Jordan}, \citenamefont {Gorshkov},\ and\ \citenamefont {Monroe}}]{Pagano_2020}%
  \BibitemOpen
  \bibfield  {author} {\bibinfo {author} {\bibfnamefont {G.}~\bibnamefont {Pagano}}, \bibinfo {author} {\bibfnamefont {A.}~\bibnamefont {Bapat}}, \bibinfo {author} {\bibfnamefont {P.}~\bibnamefont {Becker}}, \bibinfo {author} {\bibfnamefont {K.~S.}\ \bibnamefont {Collins}}, \bibinfo {author} {\bibfnamefont {A.}~\bibnamefont {De}}, \bibinfo {author} {\bibfnamefont {P.~W.}\ \bibnamefont {Hess}}, \bibinfo {author} {\bibfnamefont {H.~B.}\ \bibnamefont {Kaplan}}, \bibinfo {author} {\bibfnamefont {A.}~\bibnamefont {Kyprianidis}}, \bibinfo {author} {\bibfnamefont {W.~L.}\ \bibnamefont {Tan}}, \bibinfo {author} {\bibfnamefont {C.}~\bibnamefont {Baldwin}}, \bibinfo {author} {\bibfnamefont {L.~T.}\ \bibnamefont {Brady}}, \bibinfo {author} {\bibfnamefont {A.}~\bibnamefont {Deshpande}}, \bibinfo {author} {\bibfnamefont {F.}~\bibnamefont {Liu}}, \bibinfo {author} {\bibfnamefont {S.}~\bibnamefont {Jordan}}, \bibinfo {author} {\bibfnamefont {A.~V.}\ \bibnamefont {Gorshkov}},\ and\ \bibinfo {author} {\bibfnamefont
  {C.}~\bibnamefont {Monroe}},\ }\bibfield  {title} {\bibinfo {title} {Quantum approximate optimization of the long-range ising model with a trapped-ion quantum simulator},\ }\href {https://doi.org/10.1073/pnas.2006373117} {\bibfield  {journal} {\bibinfo  {journal} {Proceedings of the National Academy of Sciences}\ }\textbf {\bibinfo {volume} {117}},\ \bibinfo {pages} {25396–25401} (\bibinfo {year} {2020})}\BibitemShut {NoStop}%
\bibitem [{\citenamefont {Jeroslow}\ and\ \citenamefont {Wang}(1990)}]{10.1007/BF01531077}%
  \BibitemOpen
  \bibfield  {author} {\bibinfo {author} {\bibfnamefont {R.~G.}\ \bibnamefont {Jeroslow}}\ and\ \bibinfo {author} {\bibfnamefont {J.}~\bibnamefont {Wang}},\ }\bibfield  {title} {\bibinfo {title} {Solving propositional satisfiability problems},\ }\href {https://doi.org/10.1007/BF01531077} {\bibfield  {journal} {\bibinfo  {journal} {Annals of Mathematics and Artificial Intelligence}\ }\textbf {\bibinfo {volume} {1}},\ \bibinfo {pages} {167–187} (\bibinfo {year} {1990})}\BibitemShut {NoStop}%
\bibitem [{\citenamefont {PITTEL}\ and\ \citenamefont {SORKIN}(2016)}]{PITTEL_SORKIN_2016}%
  \BibitemOpen
  \bibfield  {author} {\bibinfo {author} {\bibfnamefont {B.}~\bibnamefont {PITTEL}}\ and\ \bibinfo {author} {\bibfnamefont {G.~B.}\ \bibnamefont {SORKIN}},\ }\bibfield  {title} {\bibinfo {title} {The satisfiability threshold for k-xorsat},\ }\href {https://doi.org/10.1017/S0963548315000097} {\bibfield  {journal} {\bibinfo  {journal} {Combinatorics, Probability and Computing}\ }\textbf {\bibinfo {volume} {25}},\ \bibinfo {pages} {236–268} (\bibinfo {year} {2016})}\BibitemShut {NoStop}%
\bibitem [{\citenamefont {Achlioptas}\ and\ \citenamefont {Peres}(2003)}]{Achlioptas2003}%
  \BibitemOpen
  \bibfield  {author} {\bibinfo {author} {\bibfnamefont {D.}~\bibnamefont {Achlioptas}}\ and\ \bibinfo {author} {\bibfnamefont {Y.}~\bibnamefont {Peres}},\ }\href {https://arxiv.org/abs/cs/0305009} {\bibinfo {title} {The threshold for random k-sat is $2^k ln2 - o(k)$}} (\bibinfo {year} {2003}),\ \Eprint {https://arxiv.org/abs/cs/0305009} {arXiv:cs/0305009 [cs.CC]} \BibitemShut {NoStop}%
\bibitem [{\citenamefont {Wang}\ \emph {et~al.}(2018)\citenamefont {Wang}, \citenamefont {Hadfield}, \citenamefont {Jiang},\ and\ \citenamefont {Rieffel}}]{wang2018quantum}%
  \BibitemOpen
  \bibfield  {author} {\bibinfo {author} {\bibfnamefont {Z.}~\bibnamefont {Wang}}, \bibinfo {author} {\bibfnamefont {S.}~\bibnamefont {Hadfield}}, \bibinfo {author} {\bibfnamefont {Z.}~\bibnamefont {Jiang}},\ and\ \bibinfo {author} {\bibfnamefont {E.~G.}\ \bibnamefont {Rieffel}},\ }\bibfield  {title} {\bibinfo {title} {Quantum approximate optimization algorithm for maxcut: A fermionic view},\ }\href@noop {} {\bibfield  {journal} {\bibinfo  {journal} {Physical Review A}\ }\textbf {\bibinfo {volume} {97}},\ \bibinfo {pages} {022304} (\bibinfo {year} {2018})}\BibitemShut {NoStop}%
\end{thebibliography}%

\begin{appendix}

\section{Logical Inference}
\label{app:logical-inference}

Our goal with logical inference will be to determine
\begin{enumerate}
    \item Which variables can be fixed to a specific value.
    \item Which pairs of variables are always correlated or anti-correlated.
    \item Which terms still in the Hamiltonian can be fixed to a specific value.
\end{enumerate}

To perform logical inference, we first need to restrict ourselves to one of the disconnected sub-hypergraphs formed by the constraints.  Once in this sub-hypergraph, we can find all the strings that satisfy the constraints.  This is not an efficient process, but as discussed, we expect these sub-hypergraphs to be small relative to the full problem size, with the exponential runtime just being in the size of the sub-hypergraph.  Additionally, this process can be made easier by keeping information from one iteration of the overall iterative quantum algorithm to the next, only altering things when there is a change.  

In representing these satisfying strings, we will choose to use $\pm 1$ variables, but this choice is just for representation and does not need to correspond to the logical structure of the problem being considered.  We can take these satisfying strings and set them as the rows in a matrix, $M$.  The matrix will have $n'$ columns, corresponding to the number of variables in the sub-hypergraph, and $m'$ rows, corresponding to the number of satisfying strings.  

Our logicial inference goals above can be rephrased in terms of this matrix as
\begin{enumerate}
    \item Which columns have only $+1$ or only $-1$.
    \item Which columns are exactly correlated or anti-correlated.
    \item Which terms when evaluated on every column always give a definitive and consistent value.
\end{enumerate}

The last point can be answered by running through each of the $m$ terms in the problem and evaluating them for each of the $m'$ rows, a process that will take $\mathcal{O}(m\,m'\,n')$ time and can often be shortened since many of the terms will evaluate to indeterminate or inconsistent values quickly.

The first two questions can be answered by looking at sums of columns:
\begin{enumerate}
    \item Find $h_j = \sum_{i=1}^{m'}M_{ij}\ \ \forall j\in[1,\ldots,n']$.
    \item Every $h_j$ that equals $\pm m'$ implies that the corresponding variable should be set to $\pm 1$.
    \item For all other columns make a list of candidate pairs $(r,s)$ such that $h_r=\pm h_s$ (this is a necessary but not sufficient condition for (anti-)correlation).
    \item For each $(r,s)$ pair, calculate $c_{rs} = \sum_{i=1}^{m'} M_{ir} M_{is}$.
    \item If $c_{rs}=\pm m'$ the variables corresponding to those columns should be (anti-)correlated.
\end{enumerate}

These methods will let us carry out all the logical inference we need.  In the worst case, all $h_j$ are not $\pm m'$ and are the same absolute value.  In that worst case, this algorithm takes $\mathcal{O}(n'm'+n'^2 m') = \mathcal{O}(n'^2 m')$ times.  In the best case, all the $h_j$ are $\pm m'$, in which case we get $\Omega(n'm')$ time.  We can further shortcut this algorithm if $m'>2^{n'-1}$ in which case, the pigeon-hole principle tells us that we cannot have any columns that are (anti-)correlated.

\section{Higher Locality QAOA $p=1$ Expectation Values}
\label{app:higher-locality}

We can also consider higher locality correlators when doing our analytics from Section \ref{sec:selection}.  To align with our algorithm design we consider a higher locality expectation value
\begin{equation}
    J_{Q_\alpha} = \avg{\bigotimes_{j\in Q_\alpha} \sigma_j^{(z)}}_{QAOA}
\end{equation}
where $Q_\alpha$ is again the set of nodes in the graph contained in the term in the Hamiltonian labeled by $\alpha$.  This is again considering terms that are products of Paulis.  Logical terms that are made up of a sum of products of Paulis can still be accounted for by this section because the expectation value is a linear operation.  For the rest of this section, we consider just the simple term, not summed terms.

As a further reminder, we refer to the number of qubits in this term as $k_\alpha$, and its weight in the Hamiltonian is $s_\alpha$.
The expression for $J_{Q_\alpha}$ is similar to $J_i$ above but with elements for all the qubits in the term
\begin{align}
    J_{Q_\alpha}(\beta,\gamma) &= \frac{1}{2^n}\sum_{z\in\{-1,1\}^n} 
    \sum_{
    \substack{z'_j=\pm z_j\\j\in Q_\alpha}
    }
    e^{i\gamma (C(z')-C(z))}\\\nonumber
    &
    \times\prod_{j\in Q_\alpha}\frac{1}{2}\left(e^{2i\beta}(-1)^{\frac{1+z_j}{2}} + 
            e^{-2i\beta}(-1)^{\frac{1+z'_j}{2}}\right).
\end{align}
Here $z'$ is the same as $z$ except possibly at the bits that are labeled in $Q_\alpha$, and the second summation goes over whether these bits are the same or different between $z$ and $z'$.

The first and easiest thing to do is to sum over all qubits that are not in the neighborhood of the $\alpha$ hyperedge, meaning qubits that do not share a hyperedge with any of the qubits labeled in $Q_{\alpha}$.  We will refer to this neighborhood as $N(\alpha)$ with size $|N(\alpha)| = d_\alpha$.  This neighborhood excludes the qubits in the $\alpha$ hyperedge, meaning that we can freely sum over all other bits, totaling $n-d_\alpha - k_\alpha$.

\begin{align}
    J_{Q_\alpha}(\beta,\gamma) &= \frac{1}{2^{d_\alpha+2*k_\alpha}}
    \sum_{\substack{z_k=\pm 1\\k\in N(\alpha)}}
    \sum_{\substack{z_j=\pm 1\\j\in Q_\alpha}}
    \sum_{\substack{z'_j=\pm z_j\\j\in Q_\alpha}}
    e^{i\gamma (C_\alpha(z')-C_\alpha(z))}\\\nonumber
    &
    \times\prod_{j\in Q_\alpha}\left(e^{2i\beta}(-1)^{\frac{1+z_j}{2}} + 
            e^{-2i\beta}(-1)^{\frac{1+z'_j}{2}}\right),
\end{align}
where $C_\alpha(z)$ contains all parts of the cost function that contain at least one term that contains a bit from $Q_\alpha$.

There is the possibility to simplify this expression further, but at worst, this term is now calculable numerically in time $\mathcal{O}(2^{d_\alpha+2 k_\alpha})$.

To go further than this, we need to consider all possible subsets $q\subseteq Q_\alpha$ of qubits in the term of interest.  Then $C_q(z)$ refers to all terms of the cost function (here we explicitly mean a term as a product of $\sigma^{(z)}$, not a logical term) that contain an \textbf{odd} number of bits from $q$.  Similarly, we can define $k_q = |q|$ and the neighborhood around $q$, called $N(q)$ with $d_q = |N(q)|$.

\begin{align}
    J_{Q_\alpha}(\beta,\gamma) &= 
    \sum_q
    \frac{1}{2^{d_q+k_q}}
    \sum_{\substack{z_k=\pm 1\\k\in N(q)}}
    \sum_{\substack{z_j=\pm 1\\j\in q}}
    e^{-2i\gamma C_q(z)}\\\nonumber
    &
    \times
    \cos^{k_\alpha-k_q}(2\beta)
    i^{k_q}\sin^{k_q}(2\beta)
    \prod_{i\in Q_{\alpha}}(-1)^{\frac{1+z_i}{2}}.
\end{align}
Note that $q\bigcap Q_{\alpha} \subseteq N(q)$, meaning that we are still summing over the values of all the original bits from the term.

\section{Restriction to 2-local}

\label{app:2-local}

In this appendix, we take the expression in Eq.~(\ref{eq:J_p=1}) for the $p=1$ expectation values of terms in a Hamiltonian and simplify it in the case of 2-local problems with connectivity matrices $J_{ij}$.  In that case,
\begin{equation}
    C_j(z) = \sum_{i\in N(j)}J_{ij} z_i.
\end{equation}
Then we can use the angle addition formulas for sines and cosines.  As an example, we can look at the cosines
\begin{align}
    \sum_{\substack{z_k=\pm 1\\k\in N(j)}} &\cos\left(2\gamma C_j(z)\right)=\sum_{\substack{z_k=\pm 1\\k\in N(j)}} \sum_{\text{odd}~q\geq 1}
    (-1)^{(q-1)/2}\\\nonumber
    &
    \times\sum_{A\subseteq N(j); |A|=q}\left(\prod_{i\in A}\sin 2\gamma J_{ij} z_i\prod_{i\notin A}\cos 2\gamma J_{ij} z_i\right).
\end{align}
But here, the outer summation over $z_k=\pm1$ can be moved inward explicitly because all the bits that are being summed over have decoupled.  This step is important because this would not be as easily possible with 3-local or higher terms.  In those higher locality settings, these sums need to be propogated through, considering all the information of the hypergraph formed by the neighborhood of node $j$.  But in this 2-local setting, that hypergraph is just a collection of disconnected points, making the summations easy to handle.

As we move the summations in, we can further simplify by realizing that sine is antisymmetric, so the sums over $z_i=\pm1$ will result in these sine terms evaluating to zero.  The cosines are easy to evaluate as well and evaluate to non-zero.  This insight allows us to realize that only $q=0$ can contribute.  Therefore, this entire expression evaluates to zero.

The sine terms can be evaluated with similar arguments to give that 
\begin{align}
    \sum_{\substack{z_k=\pm 1\\k\in N(j)}} &\sin\left(2\gamma C_j(z)\right)= 2^{d_j}\prod_{k\in N(j)} \cos(2\gamma J_{ij}).
\end{align}
Therefore, the form of this expression for 2-local problems is simply

\begin{align}
    J_j(\beta,\gamma) = -\sin(2\beta)
    \cos(2\gamma c_j)\prod_{k\in N(j)} \cos(2\gamma J_{ij}).
\end{align}
This form recovers known results from the literature for 2-local Hamiltonians \cite{wang2018quantum}.

\end{appendix}

\end{document}